\documentclass[a4paper,british]{article} 

\usepackage[utf8]{inputenc}

\usepackage[textwidth=16cm, textheight=23cm]{geometry}

\usepackage{multirow,bigstrut}
\usepackage{amsmath}
\usepackage{nicefrac}
\usepackage{amssymb}
\usepackage{bm}
\usepackage{dsfont}
\usepackage{authblk}
\usepackage{placeins}

\usepackage[toc,page,titletoc]{appendix}

\usepackage{soul}
\usepackage[dvipsnames]{xcolor}
\usepackage{graphicx}
\usepackage{tikz}
\usepackage{pgfplots}
\usepackage{xfrac}
\usepackage{siunitx} 
\usepackage{makecell}
\usepackage{comment}
\usepackage[bottom]{footmisc}
\usepackage{booktabs}
\usepackage{braket}
\usepackage{xspace}

\usepackage{algorithm}
\usepackage{algpseudocode}

\usepackage{caption}
\usepackage{subcaption}

\usepackage{cite}

\usepackage{hyperref}
\hypersetup{
    linkbordercolor=green,
    pdftitle={Notes_Stefan_Rothe
    },
    pdfauthor={Stefan Rothe},
    bookmarksnumbered=true,
}

\author[1]{Manoah~van~der~Worp}
\author[2]{Jonas~Krimmer}
\author[1]{Ivo~M.~Vellekoop}
\author[*,1]{Pepijn~W.~H.~Pinkse}
\author[1,2]{Stefan~Rothe}

\affil[1]{ MESA+ Institute for Nanotechnology, University of Twente, P.O. Box 217, 7500 AE Enschede, The Netherlands. }
\affil[2]{ Institute of Photonics and Quantum Electronics (IPQ), Karlsruhe Institute of Technology (KIT), Karlsruhe, Germany.}

\affil[*]{\rm{Corresponding author:} p.w.h.pinkse@utwente.nl}
\date{}                     
\newcommand{\todo}[1]{\xspace{\textcolor{red}{\bfseries[TODO: #1]}}\xspace}

\title{
Reconfigurable Linear Optical Transformations in a Single Integrated Multimode Waveguide
}

\begin{document}

\maketitle

\begin{abstract} 
Wavefront shaping enables control over optical fields for applications ranging from imaging to photonic information processing.
While conventional wavefront shaping relies on free-space systems comprising bulk optical components, compact and fully integrated approaches are comparatively unexplored.
Here, we introduce a concept based on a single multimode waveguide with distributed thermo-optic perturbations for controlling multimode propagation.
We develop a differentiable physical model and numerically demonstrate shaping the outgoing wavefront and programmable linear optical transformations, using gradient-based optimization.
We achieve diffraction-limited focusing with 20 spatial modes and show that the theoretical focusing limit can be retained while maintaining more than $99\%$ of the input optical power.
We extend this approach to complete input--output transformations and implement Sylvester--Hadamard matrix operations with dimensions up to $32\times32$ with correlations exceeding $99\%$. 
Reducing the number of tunable elements requires stronger individual perturbations, resulting in stronger and broader mode coupling that extends beyond the target modes.
To mitigate this effect, we consider ancillary modes and realize a $4\times4$ transformation that reaches $93.8\%\pm4.8\%$ correlation with only $5N^2$ tunable elements, where $N=4$ denotes the number of spatial modes. 
The best-performing realization reaches $98.3\%$ correlation with the target.
Our results demonstrate that thermo-optic control of multimode waveguides offers an approach to fully-integrated wavefront shaping and programmable linear optical transformations, with potential applications in classical and quantum photonic processing.

\end{abstract}

\section{Introduction}

The ability to control the spatial wavefront of light is crucial for a wide range of optical technologies, including high-resolution microscopy and imaging~\cite{horstmeyer2015guidestar,kubby2019wavefront,maurer2011spatial,booth2007adaptive,park2018perspective, gigan2022roadmap,stellinga2021time,cao2023controlling,bertolotti2022imaging,gomes2025funnelling}, secure optical communications~\cite{goorden2014quantum,bromberg2019remote,rademacher202010,rothe2023securing,amitonova2020quantum}, 
multimode fiber lasers and amplifiers~\cite{florentin2017shaping,rothe2025wavefront,lee2026reconfigurable,fu2018several,cao2023spatiotemporal,wei2020harnessing}, 
high-dimensional quantum information processing~\cite{defienne2016two,wolterink2016programmable,leedumrongwatthanakun2020programmable,valencia2020unscrambling,goel2026quantum}, and optical processing and computing~\cite{momeni2025training,teugin2021scalable,xia2024nonlinear,wright2022deep,wang2022optical,hu2024diffractive,yildirim2024nonlinear,xu2026chip,han2026optical,xu2026chip,cheng2024multimodal,fu2023photonic,zhu2022space}. 
Most of the remarkable advances achieved in these fields have been enabled by more than two decades of intensive research in wavefront shaping. 
Beginning with the demonstration of focusing light through opaque media using phase-only modulation~\cite{vellekoop2007focusing}, the field has evolved toward increasingly sophisticated control approaches, including emphasis on light efficiency~\cite{rocha2024fast}, spatio-temporal control~\cite{aulbach2012spatiotemporal,carpenter2016complete,velsink2020spatiotemporal,cruz-delgado2022synthesis,shen2023roadmap}, and polarization control~\cite{xiong2018complete,mounaix2019control,rothe2025output}. 
Modern wavefront-shaping systems allow for control of thousands to millions of spatial degrees of freedom~\cite{Fontaine2021hermite,ammar2025upper}, and allow for simultaneous manipulation of the phase, amplitude, spectrum, polarization or their combinations of optical fields~\cite{mounaix2020time,yessenov2022vector}.
At the most general level, these capabilities are governed by the underlying optical transmission operator, which describes the mapping between the available input and output degrees of freedom, called modes~\cite{cao2022shaping}.
Numerous studies have demonstrated that system performance measures such as communication capacity, laser power instability threshold, computational complexity, and focusing enhancement improve with the number of independently controlled modes~\cite{Miller2019waves,wright2022nonlinear}. 

The practical realization of wavefront shaping typically relies on spatial light modulators~(SLMs) or digital micromirror devices~(DMDs), which provide reconfigurable access to available modes of an optical system~\cite{Leith1966Holographic,Brown1966complex,Lee1978computer, Turtaev2017Comparison}.
These devices generally require bulk free-space optics comprising lenses, mirrors, and spatial or spectral filtering elements. The resulting systems require careful alignment, occupy a comparatively large footprint, and are susceptible to mechanical and thermal drifts. 
These challenges become increasingly restrictive as the number of controlled spatial modes increases~\cite{gomes2022near,ammar2025upper}.

Photonic integrated circuits~(PICs) enable compact and ultrafast wave control\cite{cao2019reconfigurable}, as they transfer optical functionalities to the chip-scale~\cite{bogaerts2020programmable}.
Programmable PICs are relevant to a wide range of applications, including beam steering~\cite{sun2013large,heck2017highly}, all-optical mode (de-)multiplexing~\cite{lu2024empowering,nakajima2025programmable,wang2026reconfigurable}, quantum state generation and manipulation~\cite{carolan2015universal,chen2024heralded}, and demonstrations of noisy intermediate-scale quantum photonic processors~\cite{tillmann2013experimental,spring2013boson,wang2019boson,somhorst2023quantum}. 
Most PIC designs that involve manipulation of multiple modes rely on meshes of single-mode waveguides and tunable Mach--Zehnder interferometers~\cite{reck1994experimental, clements2016optimal,Taballione2019reconfigurable}.
However, scaling such architectures to large numbers of spatial channels requires increasingly large networks of nominally identical interferometric unit cells. 
Multimode waveguides~(MMWs) offer a particularly compact approach to spatial optical processing because the spatial modes co-propagate simultaneously within a single waveguide.
A suitable modulation mechanism can therefore tune the available modes collectively within the same physical structure.
Control of multimode propagation has been demonstrated in MMWs using externally programmed refractive-index modifications~\cite{bruck2016all,onodera2024scaling,yanagimoto2026programmable}, while integrated diffractive processors have demonstrated trainable multimode propagation using distributed thermo-optic perturbations~\cite{zhu2022space,fu2023photonic,cheng2024multimodal,xu2026chip}. Together, these results demonstrate the potential of controlled multimode propagation as a resource for programmable optical processing.

Here, we simulate the design of thermally tunable MMWs that is a particularly simple integrated architecture: the programmable optical processor consists of a single MMW with distributed thermo-optic control and allows to directly program the transformation of the MMW itself.
We position thermo-optically tunable elements along the waveguide and generate transverse refractive-index perturbations that couple the guided modes, while propagation between successive perturbations introduces mode-dependent phase accumulation.
Their collective action enables control over the input--output transformation within the MMW. 
Our concept is inspired by perturbation-based wavefront shaping in large-scale multimode fibers, where controlled refractive-index perturbations have been used to engineer the effective transmission matrix of complex optical systems~\cite{resisi2020wavefront,finkelstein2023spectral,shekel2024tutorial}.
While the thermo-optic phase shift of a single-mode waveguide is well understood~\cite{jacques2019optimization}, 
a corresponding physical description relating distributed thermo-optic perturbations to the complete complex input--output transformation of an MMW, however, is comparatively unexplored.
We therefore develop a physical model that directly connects the local thermo-optic perturbations to the propagating electromagnetic field and the resulting optical transformation.
Although propagation remains linear in the input field, its dependence on the control parameters is nonlinear, motivating multivariable optimization. 
Using a differentiable forward model and gradient-based optimization, we demonstrate diffraction-limited focusing as well as programmable optical transformations with dimensions up to $32\times32$. 

\section{Thermally perturbed multimode waveguides}

\begin{figure}[h]
    \centering
    (a)\includegraphics[width=0.94\textwidth]{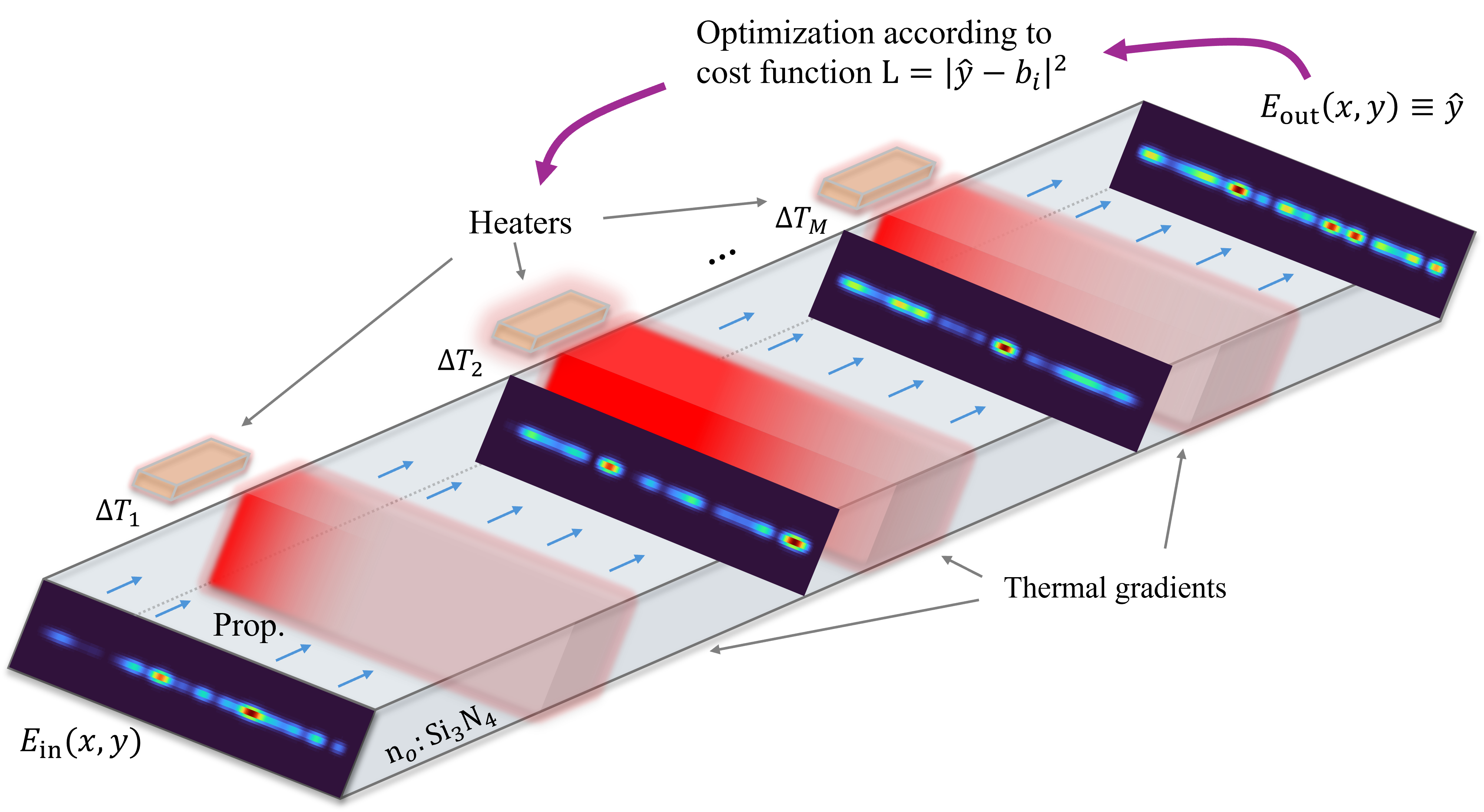}\\

    (b) \includegraphics[width=0.67\textwidth]{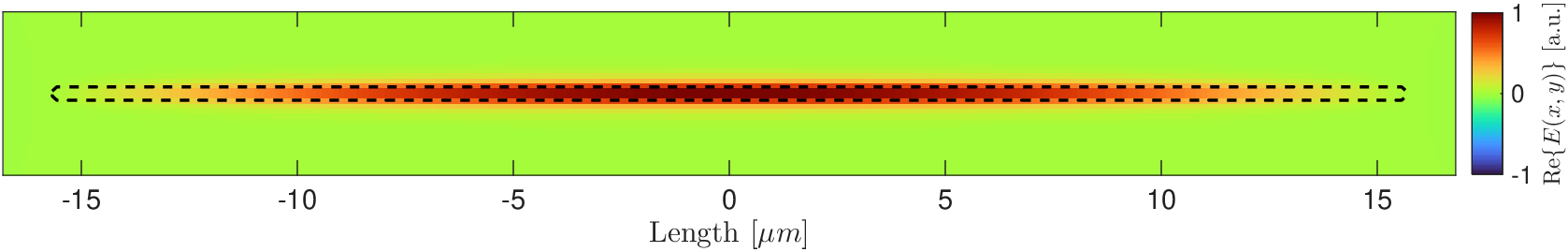}\\
    (c)
    \includegraphics[width=0.67\textwidth]{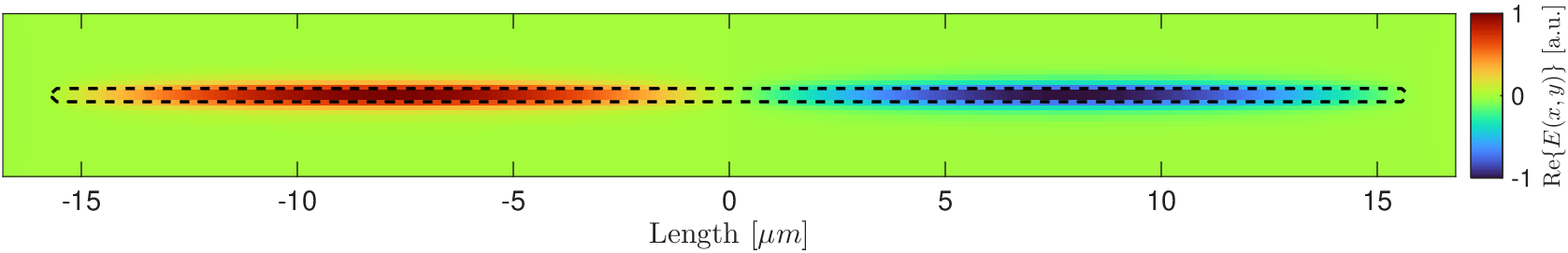}\\
    (d)
    \includegraphics[width=0.67\textwidth]{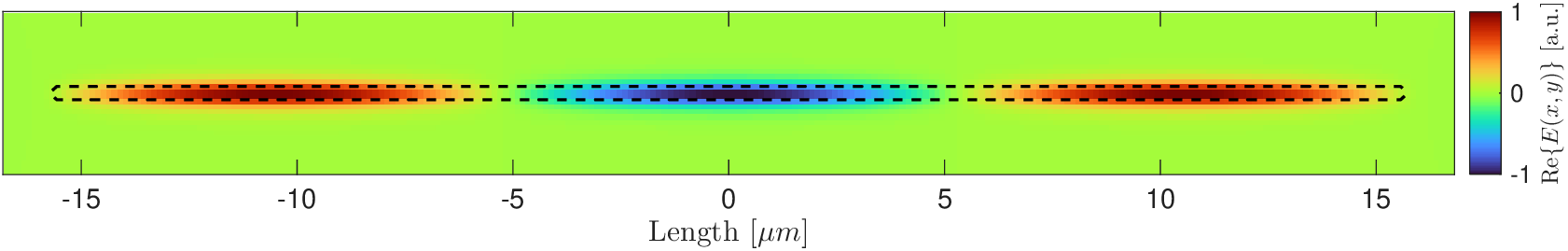}\\

\caption{ Thermally programmable MMW. (a) Schematic of the proposed device. An input field $E_{\mathrm{in}}(x,y)$ consists of an arbitrary superposition of modes and propagates through a sequence of $M$ thermo-optic control sections. Localized refractive-index perturbations induced by metallic heaters that generate temperature differences $\Delta T_i$ enable programmable mode coupling and phase modulation. The resulting output field $E_{\mathrm{out}}(x,y)$  is obtained and used as feedback for optimization, according to a cost function $L$. (b)-(d) Normalized electric-field distributions $\mathrm{Re}\{E(x,y)\}$
of the first three guided modes of an unperturbed $26$-\textmu m-wide SiN MMW at $\lambda=785~\mathrm{nm}$ wavelength. The black dashed rectangle indicates the core-cladding boundary.}
    \label{fig:1}
\end{figure}

\begin{figure}[h] 
    \centering 
(a)\includegraphics[height=4.7cm,trim=1.5cm 0 1cm 0,
    clip]{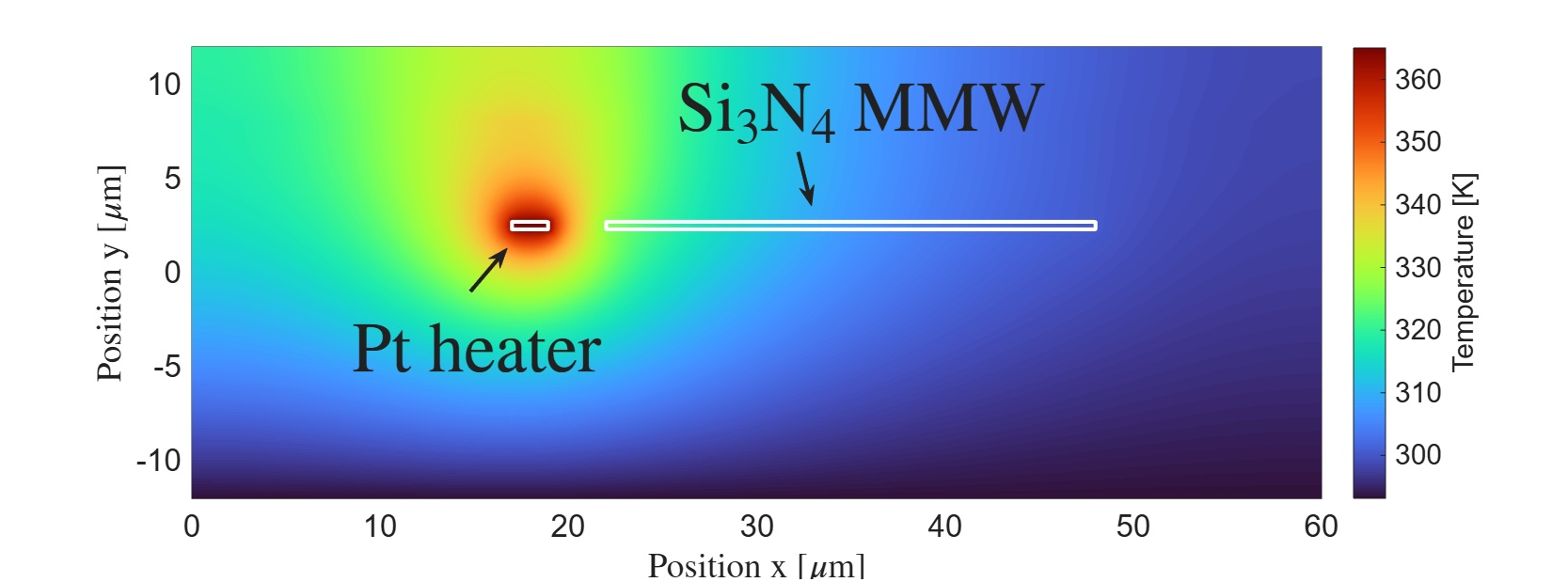} (b)\includegraphics[height=4.5cm]{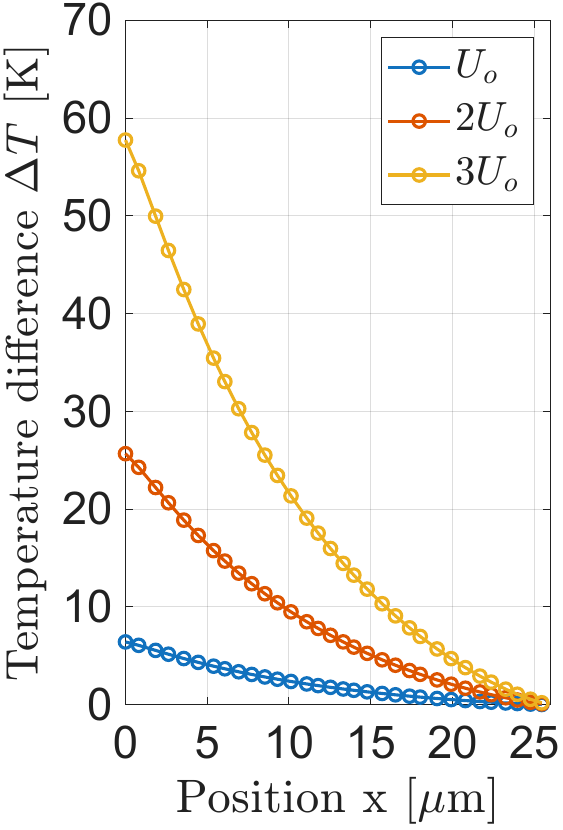}\\ \caption{ 
    Thermal control mechanism. (a) Simulated temperature distribution obtained from finite-element analysis for a Pt heater placed $3$~\textmu m adjacent to the MMW core. Joule heating produces a spatially non-uniform temperature profile across the MMW. (b) Temperature profiles across the MMW core for three different heater voltages ($U_o$, $2U_o$, and $3U_o$). The resulting thermal gradients constitute the control mechanism used to induce refractive-index perturbations and programmable mode coupling.} \label{fig:2} 
\end{figure}

The proposed architecture is based on thermally programmable silicon nitride ($\mathrm{Si}_3\mathrm{N}_4$, SiN) MMWs, as illustrated in Fig.~\ref{fig:1}a. 
Throughout this work, we consider a $26$-\textmu m-wide SiN MMW surrounded by a $\mathrm{SiO}_2$ cladding, supporting $52$ guided spatial modes at a wavelength of $785$~nm.
Figure~\ref{fig:1}b shows the first three guided modes of a representative MMW, i.e., $\mathrm{Re}\{E(x,y)\}$. 
Details of the mode calculations are provided in the 
Supplementary material.
We consider an arbitrary superposition of guided spatial modes $E_{\mathrm{in}}(x,y)$ that is excited at the input facet and propagates through the waveguide. 
The resulting output field $E_{\mathrm{out}}(x,y)$ serves as the feedback signal $\hat{y}$ for the optimization procedures introduced in subsequent sections.
At selected longitudinal positions, integrated metallic heaters are placed in close distance to the MMW. 
The dissipated electrical power generates transverse temperature gradients $\Delta T(x,y)$ across the waveguide cross-section, which translate into spatial refractive-index perturbations through the thermo-optic effect. 
The underlying thermal mechanism is illustrated in Fig.~\ref{fig:2}a. 
A metallic heater that is positioned $3$~\textmu m adjacent to the $26$\textmu m-wide MMW core generates localized Joule heating~\cite{jacques2019optimization}, resulting in a spatially non-uniform temperature distribution.
The temperature distributions shown are obtained from finite-element simulations performed in COMSOL Multiphysics, with details provided in the 
Supplementary Material. 
Figure~\ref{fig:2}b shows the corresponding temperature profiles $\Delta T(x)$ across the MMW core for three different voltages applied to the heater.
We approximate the resulting temperature profiles across the MMW core by a linear transverse gradient.
The slope of the gradient determines the strength of the thermo-optic perturbation and can be adjusted through the electrical power dissipated in the heater.
For the strongest perturbation considered (yellow curve in
Fig.~\ref{fig:2}b), the root-mean-square deviation from a linear fit,
normalized to the total temperature variation across the MMW core, is
$7.5\%$. 

A representative three-dimensional heater extends over several millimeters along the propagation direction to induce the required thermo-optic phase modulation.
Three-dimensional thermal simulations of such a heater are provided in Fig.~\ref{fig:heater3D} of the Supplementary Material.
As shown in Fig.~\ref{fig:heater3D}b, the temperature distribution is nearly uniform along the longitudinal direction, with variations occurring predominantly near its ends.
We therefore approximate each heated section by a longitudinally uniform thermo-optic perturbation and represent its accumulated effect by a two-dimensional phase screen.

For a monochromatic wave propagating through a heated section of length $\delta L$, the accumulated phase shift is given by 
\begin{equation}
\Delta\phi (x,y) = k_0\, \frac{dn}{dT}\,\Delta T (x,y)\,\delta L,
\label{eq:phase-shift}
\end{equation}
where $k_0=\frac{2\pi}{\lambda_0}$ and $\frac{dn}{dT}$ denotes the thermo-optic coefficient~\cite{jacques2019optimization}. 
In thermally perturbed single-mode waveguides, the effect of $\Delta\phi(x,y)$ can be absorbed into a change of the propagation constant, such that the thermo-optic control results primarily in an accumulated phase shift. 
In an MMW, however, the spatial variation of $\Delta\phi(x,y)$ across the core must be accounted for, since the resulting refractive-index perturbation couples the transverse modal eigenstates. 
Between successive perturbation sections, the modes accumulate additional relative phases due to modal dispersion. 
By cascading $M$ such perturbation-and-propagation sections, the optical field undergoes a sequence of controlled amplitude and phase modulation events. 
Through appropriate heater settings, not only the output field $E_{\mathrm{out}}(x,y)$, but also the entire input-output transmission operator can therefore be reconfigured. 
In the following section, we develop a forward-model that describes these multimode interactions and establishes the relation between the heater settings and the resulting output field.

\section{Thermo-optic mode control and optimization}
\label{sec:model}

The remaining challenge is to determine the optimal heater settings required to generate a target output field. 
While the heater temperatures, and therefore the induced refractive-index perturbations, can be directly controlled through the applied voltages, i.e. $\Delta \phi(x,y) \propto \Delta T(x,y) \propto \Delta U$, the resulting dependence of the output field on these control parameters is generally nontrivial.

Here, we employ a differentiable forward model describing multimode propagation through the MMW. The model is formulated in an orthonormal modal basis obtained from numerical eigenmode simulations of the unperturbed SiN waveguide. Details of the MMW simulation parameters are provided in the Supplementary Material. 
For a given wavelength, the eigenmode solver provides the transverse mode profiles $\psi_j(x,y)$ and propagation constants $\beta_j$ of the guided modes, which are shown in Fig.~\ref{fig:modeSet}a of the Supplementary Material. 
The guided modes form an orthonormal basis, as verified by their mutual modal overlaps shown in Fig.~\ref{fig:modeSet}b of the Supplementary Material.
The optical field is represented as $ E(x,y,z) = \sum_{j=1}^{N} a_j(z) \, \psi_j(x,y)\, e^{i\beta_j z}$, where $a_j(z)$ denotes the complex modal weight of mode $j$, and $N$ is the number of guided modes included in the simulation. 
Propagation over the longitudinal distance $z_k$ is described by phase accumulation according to the modal propagation constants $e^{i\beta_j z_k}$. Here, the distance $z_k$ is defined as the spacing between two subsequent heaters. 
At each heater position $z+z_k$, the modal field is reconstructed in the
spatial domain and perturbed by a phase screen induced by the thermo-optic
refractive-index modulation,
\begin{equation}
    E_{\mathrm{pert}}(x,y,z_{k})
    =
    E(x,y,z_{k})\,
    e^{i\Delta \phi_k(x,y)},
    \label{eq:perturbation}
\end{equation}
where $\Delta \phi_k(x,y)$ denotes the phase profile induced by the $k$th heater. 
As shown in Fig.~\ref{fig:2}b, the thermo-optic perturbation generates an approximately linear transverse temperature gradient across the MMW core, which translates into a corresponding phase modulation. 
The spatially varying phase perturbation modifies the transverse field distribution and thereby couples the guided modes. 
Figure~\ref{fig:thermalCoupling} of the Supplementary Information illustrates the resulting coupling among the first $20$ guided modes for different perturbation strengths. 
As the perturbation strength increases, the modal coupling becomes progressively stronger, extends over a broader range of modes and occurs predominantly between neighboring modes.
Importantly, the resulting intensity response is intrinsically nonlinear with respect to the perturbation strength.
This behavior is illustrated in Fig.~\ref{fig:single_heater_response} of the Supplementary Information, where the perturbation strength $\xi_k$ of individual tunable elements is varied while all remaining perturbations are set to zero.
For small variations around $\xi_k=0$, the intensity within a fixed target region at the MMW output exhibits an approximately linear dependence on $\xi_k$.
As the perturbation strength increases, the response progressively deviates from this local linear behavior and becomes increasingly nonlinear.

For simplicity, we approximate the two-dimensional phase perturbation induced by each heater by a one-dimensional linear phase gradient across the MMW core.
Since the MMW is single-mode in the $y$-direction, an additional variation along $y$ is not required.
The phase profile of the $k$th tunable element is therefore described by
\begin{equation}
    \Delta\phi_k(x;\xi_k)
    =
    \xi_k\,\frac{2\pi}{W}x + \phi_{0,k},
    \label{eq:phase_gradient}
\end{equation}
where $W$ denotes the width of the MMW and $\phi_{0,k}$ is a spatially uniform phase offset.
The dimensionless parameter $\xi_k$ fully determines the transverse phase gradient and thus the strength of the thermo-optic perturbation.
A value of $|\xi_k|=1$ corresponds to a total phase variation of $2\pi$ across the MMW core.
As explained in the next section, the heaters are placed alternately on opposite sides of the MMW, such that the direction of the induced transverse phase gradient can change between successive tunable elements.
This alternating gradient direction is explicitly accounted for in the simulations and is described in detail in the Supplementary Information.


The set of phase-gradient slopes $\boldsymbol{\xi}=\{\xi_1,\xi_2,\dots,\xi_M\}$ constitutes the control parameters of the system and is optimized for all optimizations considered in this work. 
Because the perturbation modifies the local refractive-index profile, the unperturbed modes no longer represent the exact propagation eigenstates within the heated section. 
The resulting mode coupling is obtained by projecting the perturbed field onto the modal basis of the unperturbed waveguide,
\begin{equation}
    a_j'
    =
    \iint
    \psi_j^*(x,y)\,
    E_{\mathrm{pert}}(x,y,z_{k})\,
    dx\,dy,
    \label{eq:modal_decomp}
\end{equation}
which yields updated modal coefficients $a_m'$ after each thermo-optic perturbation. 
This procedure is repeated sequentially for all thermo-optically tunable elements along the propagation direction. 
Importantly, the projection in Eq.~(\ref{eq:modal_decomp}) does not generally conserve the optical power within the guided modes, as part of the perturbed field may couple to radiation modes and therefore leak from the waveguide, unless mitigating steps are taken. 
Such perturbation-induced loss is explicitly accounted for in the cost functions, as described in the Supplementary Material. 

After propagating through all $M$ thermo-optic control sections, we obtain the output field $E_{\mathrm{out}}(x,y)$. Depending on the target application, different cost functions for shaping $E_{\mathrm{out}}$ can be defined.
The goal of the optimization is to determine the set of the phase gradient slopes $\boldsymbol{\xi}$ that minimizes a specific cost function.
In this work, we consider wavefront-shaping tasks such as focusing light into one or multiple regions at the MMW output facet, as well as optimization of the input-output transmission matrix following a target unitary transformation. 
The corresponding cost functions are summarized in detail in the Supplementary Material. 
For a given cost function $\mathcal{L}(\xi_k)$, the gradient with respect to the control parameters $g_{j,\xi_k}=\frac{\partial \mathcal{L}_j}
{\partial \xi_k}$ is calculated in every iteration $j$. The heater settings are updated according to

\begin{equation}
 \xi_{j+1,k} = \xi_{j,k} - \alpha \, g_{j,\xi_k},   
\end{equation}

where $\alpha$ denotes the adaptive learning rate obtained through the Adam optimizer~\cite{kingma2017adammethodstochasticoptimization}. 
This procedure is repeated iteratively until convergence is reached. 
Such a gradient-descent-based optimization has recently demonstrated excellent performance in optical beam shaping problems, including nonlinear wave propagation~\cite{barre2022holographic,kupianskyi2023high,rothe2025output}. 
Such approaches are known to be powerful with nonlinear parameter dependencies and scale efficiently to systems with hundreds of controllable degrees of freedom. 
The structure of our algorithm is explained in Algorithm\ref{alg:mmw_optimization} in the Supplementary Material.

\section{Single-spot focusing}

\begin{figure}[htb]
    \centering
    (a)\includegraphics[width=0.47\textwidth,trim=4cm 0 1cm 0.5cm,
    clip]{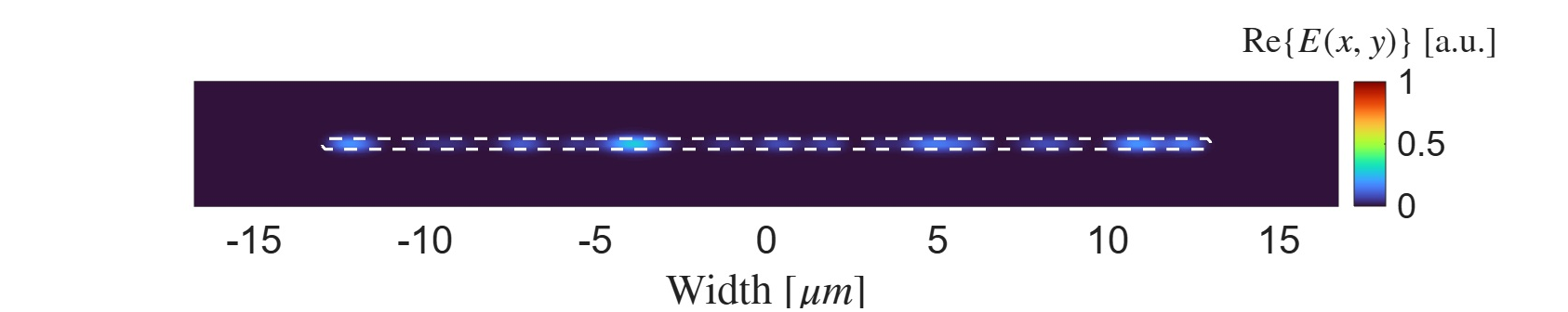}
    (b)\includegraphics[width=0.46\textwidth,trim=4cm 0 1cm 0.5cm,
    clip]{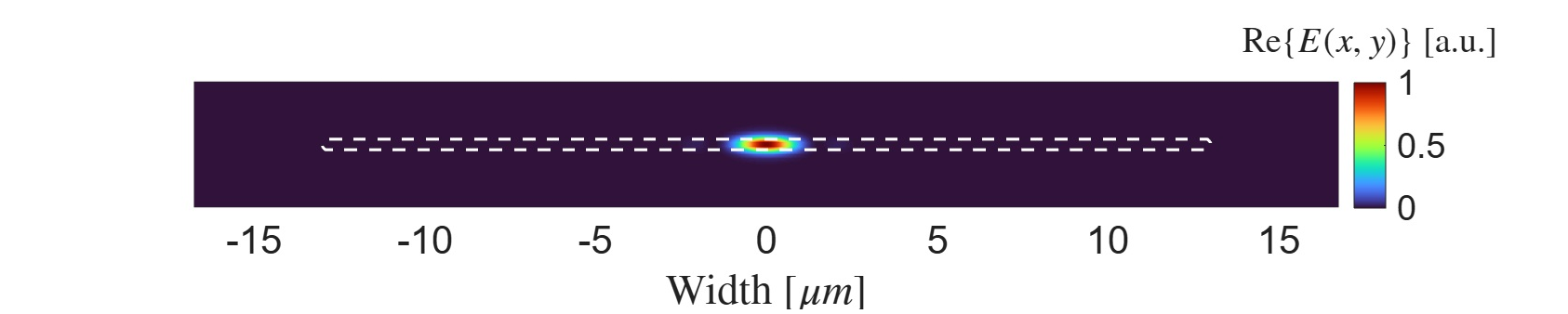}\\
    (c)\includegraphics[width=0.47\textwidth]{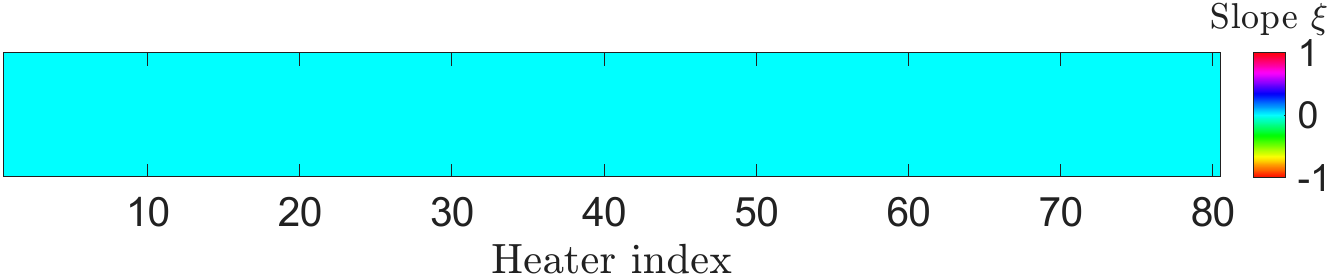}
    (d)\includegraphics[width=0.47\textwidth]{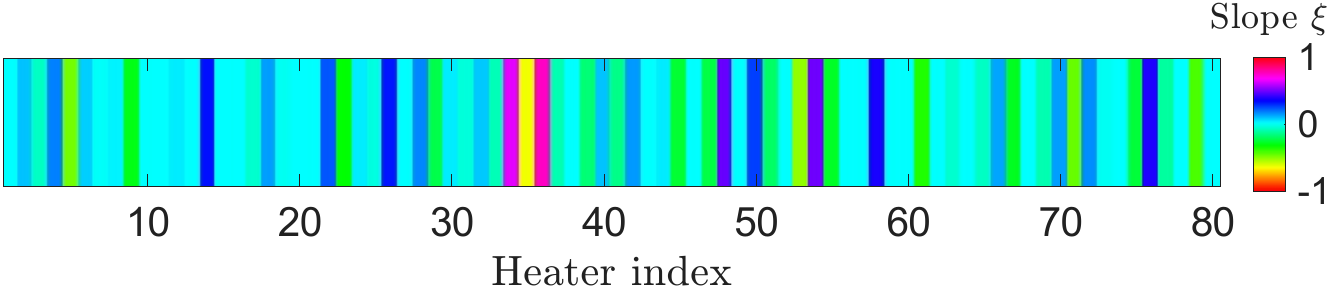}\\
    (e)\includegraphics[height=4.2cm]{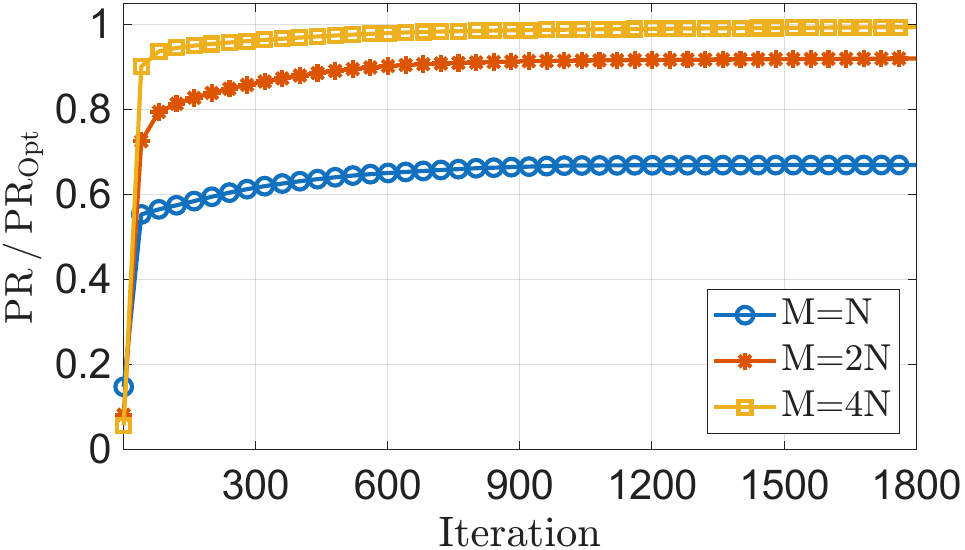}
    (f)\includegraphics[height=4.2cm]{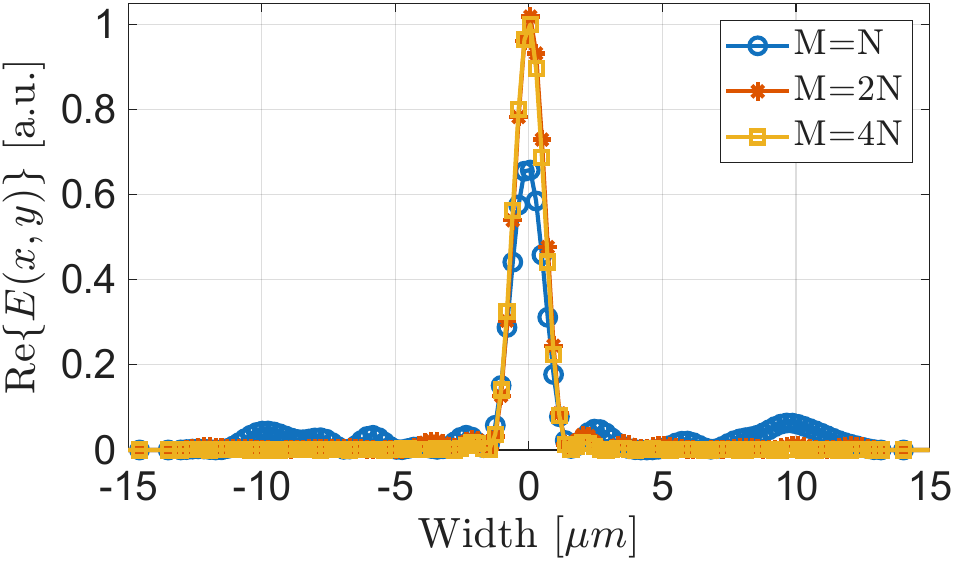}\\
    \caption{Wavefront shaping in an $26$-\textmu m-wide and 16-cm-long MMW supporting $N=52$ guided modes at $\lambda=785$~nm, using the cost function in Eq.~(\ref{eq:cost_single_spot_focusing}) with $\gamma=0$. The learning rate is set to $\alpha=0.01$. Only the first $20$ guided modes are excited with uniform optical power and random relative phases. (a) Normalized real part ofthe output field $\mathrm{Re}\{E(x,y)\}$ before and (b) after gradient-based optimization using $M=4N=80$ thermo-optically tunable elements. The white dashed circle denotes the core region 
    (c) Corresponding phase-gradient slopes $\boldsymbol{\xi}$ of
    the $M=80$ tunable elements before and (d) after optimization. 
    (e) Evolution of the PR normalized to its theoretical upper bound for $M=N$, $M=2N$, and $M=4N$ tunable elements. For $M=4N$, the optimization reaches the theoretical upper bound $\mathrm{PR_{Opt}}\approx84\%$, corresponding to a normalized $\mathrm{PR}\approx 100\%$.
    (f) Line scans through the center of the corresponding output fields, showing $\mathrm{Re}\{E\}$ normalized to the maximum field magnitude. 
    Increasing $M$ progressively suppresses the background field and concentrates more optical power within the target focal region.}
    \label{fig:3}
\end{figure}

Perhaps the most well-known demonstration of wavefront shaping is the focusing of a coherent light beam through a strongly scattering medium, where a diffraction-limited focus is formed behind an otherwise opaque sample that would otherwise produce a random speckle pattern~\cite{vellekoop2007focusing}. 
Likewise, we demonstrate the formation of a diffraction-limited focus at the output facet of the MMW.
In contrast to a scattering medium, the uncontrolled speckle-like output originates from modal dispersion.
The top panel of Fig.~\ref{fig:3}a shows the output intensity distribution resulting from uncontrolled propagation through a $26$-\textmu m-wide MMW, supporting $52$ guided modes at a wavelength of $785$~nm. 
The corresponding propagation constants are shown in Fig.~\ref{fig:modeSet}a. 
We excite the first $N=20$ guided modes with equal optical power and random relative phases. 
For simplicity, coupling into higher-order modes that are not excited at the input is neglected. 
To investigate the number of tunable elements required for intensity shaping, we consider $M=N$, $2N$, and $4N$ thermo-optically tunable elements, corresponding to $M=20$, $40$, and $80$, respectively. 
The elements are distributed along the propagation direction with $z_k=50$~\textmu m longitudinal spacing, adjacent to the waveguide core and alternating between the left- and right-hand sides, with the first heater positioned on the left-hand side. 
With this spacing and assuming an upper-bound heater length of $2$~mm, as discussed in the Supplementary Material, the MMW containing 80 tunable elements has a total optical path length of $\approx 16$~cm.
Figures~\ref{fig:3}a--d show the initial and optimized fields and control settings for $M=4N=80$.

Even over the relatively short propagation distance inside the MMW, modal dispersion causes the relative modal phases to evolve, resulting in a strongly decorrelated transverse intensity distribution at the output. 
Modal dispersion alone scrambles the field into a speckled intensity distribution at the output.

Taking Fig.~\ref{fig:3}a as the initial state of $E_{\mathrm{out}}(x,y)$, the central region of the output facet is defined as the target area $\Omega_{\mathrm{target}}$, which is defined as twice the width of the $e^{-2}$ drop from the maximum intensity of the corresponding ideal diffraction-limited focus, and is used to calculate the corresponding power ratio $\mathrm{PR}$~\cite{gomes2022near}:
\begin{equation}
    \mathrm{PR}=\frac{
\iint_{\Omega_{\mathrm{target}}}
\lvert E_{\mathrm{out}}(x,y) \rvert ^2
\,dx\,dy
}
{
\iint
\lvert E_{\mathrm{out}}(x,y) \rvert ^2
\,dx\,dy
},
\label{eq:pr}
\end{equation}

which corresponds to the fraction of the total optical power contained within the target region. 
The instantaneous $\mathrm{PR}$ enters the cost function, which is used to optimize the phase-gradient slopes, maximizing the optical power within the target region. 
A detailed description of the cost function is provided in the Supplementary Material. 
For the results shown in Fig.~\ref{fig:3}, we set $\gamma=0$, since the optimization already results in negligible optical loss without explicitly constraining the transmitted power. 
The influence of nonzero $\gamma$ on the trade-off between focusing performance and optical loss is investigated in the Supplementary Information in Fig.~\ref{fig:gamma_comparison}.

In the initial state, all heaters are switched off, such that $\boldsymbol{\xi}=\mathbf{0}$.
Figure~\ref{fig:3}c illustrates the corresponding set of phase-gradient slopes for $M=4N=80$ before optimization.
Executing the gradient-descent optimization creates a sharp focal spot, as shown in Fig.~\ref{fig:3}b.
The resulting optimized phase-gradient slopes $\boldsymbol{\xi}=\{\xi_1,\xi_2,\dots,\xi_M\}$ are shown in Fig.~\ref{fig:3}d and exhibit no obvious deterministic structure.
The optimized control distribution has a participation ratio of $\epsilon=0.41$, as defined in Eq.~(\ref{eq:participation_ratio}) in the Supplementary Material, indicating that the control is distributed over a fraction of the available tunable elements.

We next investigate how the focusing performance depends on the number of available tunable elements.
Figure~\ref{fig:3}e shows the evolution of the PR for $M=N$, $M=2N$, and $M=4N$, normalized to the theoretical upper bound, $\mathrm{PR}_{\mathrm{opt}}\approx84\%$.
Our approach for determining $\mathrm{PR}_{\mathrm{opt}}$ is described below.
Here, one iteration corresponds to a simultaneous update of all tunable elements.
For $M=N=20$, the optimization converges to an absolute PR of approximately $56\%$, corresponding to $67\%$ of $\mathrm{PR}_{\mathrm{opt}}$.
Increasing the number of tunable elements to $M=2N=40$ increases the normalized PR to approximately $93\%$ of $\mathrm{PR}_{\mathrm{opt}}$, while $M=4N=80$ reaches $100\%$ of $\mathrm{PR}_{\mathrm{opt}}$.
We repeated these simulations for different random input states and observed only negligible variations in the optimum PR, indicating that the achieved focusing performance is largely insensitive to the particular input state.

The corresponding line scans through the focal region are shown in Fig.~\ref{fig:3}f.
Here, $\mathrm{Re}\{E\}$ is normalized to the maximum field magnitude for each optimized output field.
Increasing the number of tunable elements progressively suppresses the background field surrounding the focus and concentrates a larger fraction of the optical power within the target region, consistent with the corresponding increase in PR.

For the considered thermo-optic perturbation geometry, $4N$ tunable elements are sufficient to reach the maximum focusing performance.
In general shaping the intensity of a field represented by $N$ spatial modes, $N$ independent control degrees of freedom are in principle sufficient, provided the applied degrees of control form an orthogonal basis.
The thermo-optic gradients considered here, however, do not form such an orthogonal basis.
Individual perturbations therefore provide partially redundant control over the propagating field, requiring a larger number of tunable elements.
This is reflected in the present simulations, where $M=N$ remains below the focusing limit, whereas $M=4N$ is sufficient to reach the theoretical upper bound.

The upper bound of $\mathrm{PR}$ is determined iteratively from the set of guided modes.
We first define a single bright pixel at the center of $\Omega_{\mathrm{target}}$ and project it onto the available modes, analogous to the modal projection in Eq.~(\ref{eq:modal_decomp}). 
The reconstructed field already forms a localized focus, but a fraction of the optical power remains distributed outside the focal region, resulting in a $\mathrm{PR}$ below the optimum value.
To further concentrate the field, we truncate the reconstructed focus at the $e^{-2}$ intensity boundary and project the resulting field onto the modal basis again.
This truncation and re-projection procedure progressively suppresses power outside the focal region and increases the $\mathrm{PR}$.
The procedure is repeated until the $\mathrm{PR}$ converges, which typically occurs after approximately three iterations.
The corresponding $\mathrm{PR}$ represents the maximum focusing performance achievable with the available set of guided modes and is therefore used as the theoretical upper bound.

For all three configurations shown in Fig.~\ref{fig:3}, approximately $82\%$ of the input optical power is retained after propagation through the perturbed MMW.
As discussed in the following sections, increasing the number of available control elements allows the desired optical transformation to be realized with weaker individual perturbations, thereby reducing the maximum required phase-gradient slope.
A larger number of control elements can therefore enable smoother perturbations and provide additional freedom to optimize the trade-off between focusing performance and optical loss.
For this purpose, the cost function [Eq.~\ref{eq:cost_single_spot_focusing}] additionally contains a term with weight $\gamma$ that penalizes perturbation-induced power loss.
As shown in Fig.~\ref{fig:gamma_comparison} of the Supplementary Material, using $M=400$ control elements for the same $N=20$ excited modes and $\gamma=2$ allows the theoretical focusing limit to be reached while retaining more than $99\%$ of the input optical power.

\section{Programmable matrix transformations in multimode waveguides}

\begin{figure}[t]
    \raggedright
    (a)\includegraphics[height=5cm]{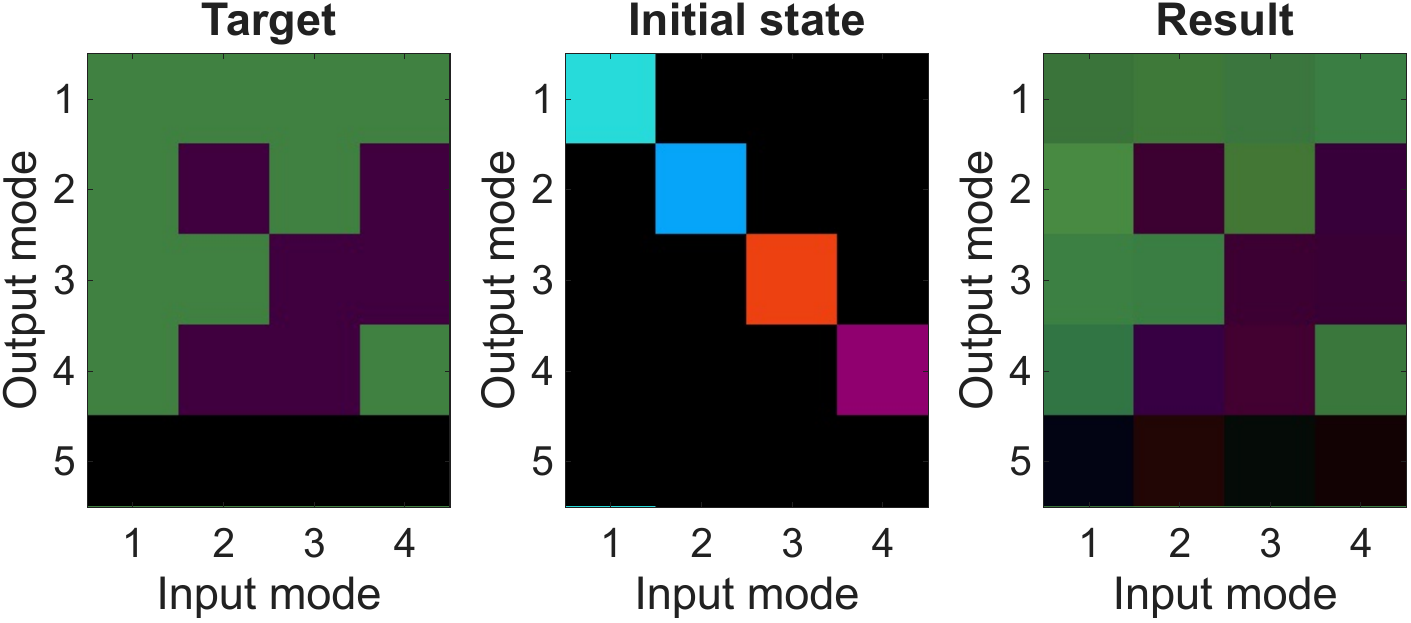} \includegraphics[height=5cm]{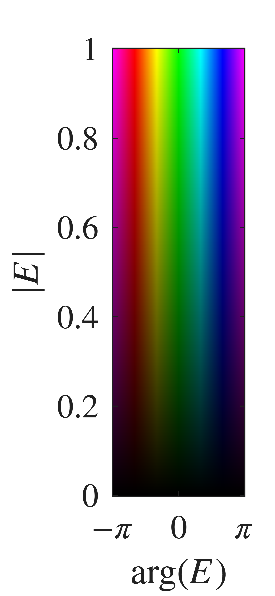}\\
    (b)\includegraphics[height=4.2cm]{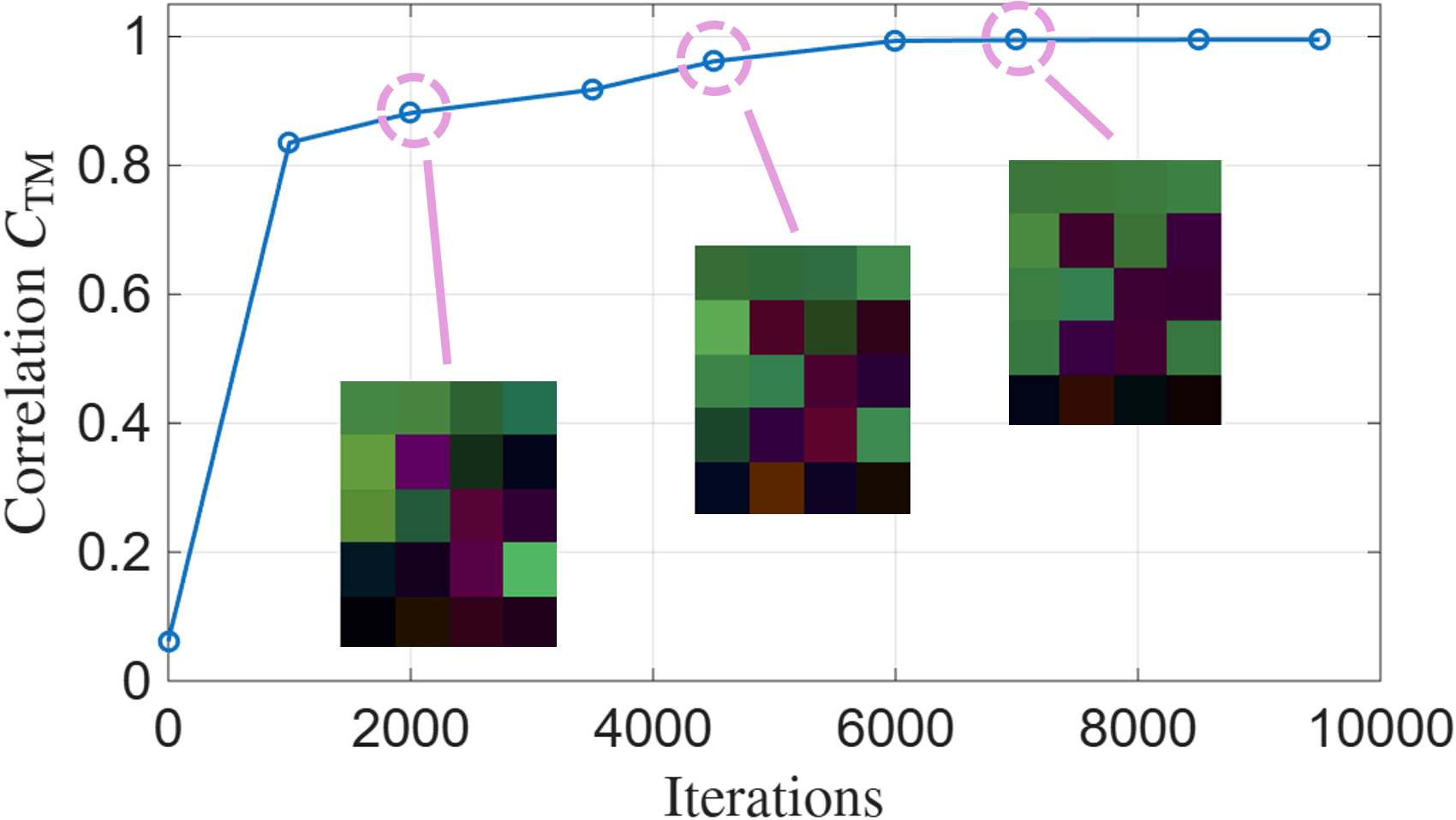}
    (c)\includegraphics[height=4.5cm]{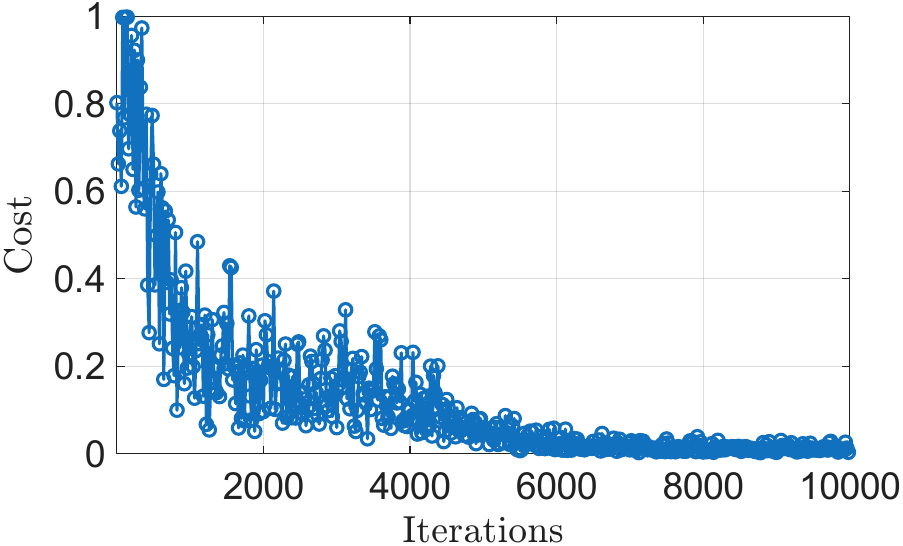}\\
    (d)\includegraphics[width=0.85\textwidth]{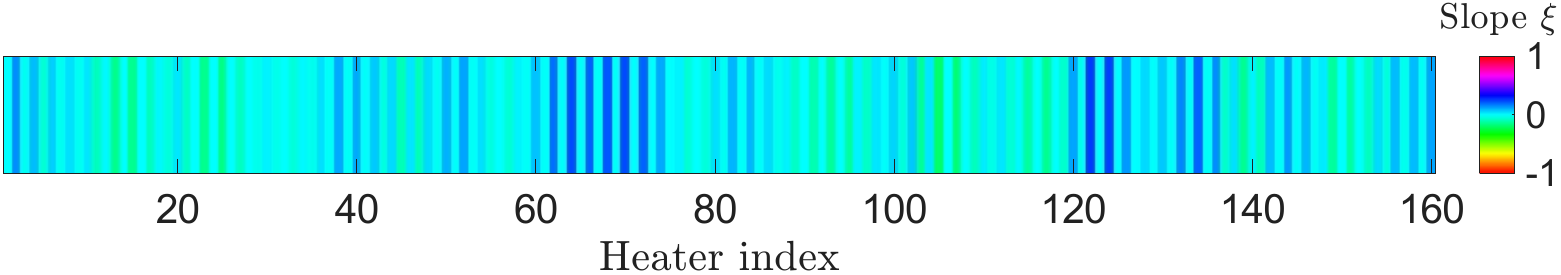}\\  
    \caption{Optimization of a $4\times4$ Sylvester--Hadamard transformation. The learning rate was set to $\alpha=2.6\times 10^{-4}$.
    (a) Target (left), initial (center), and optimized (right) transmission matrices for $N=4$ target modes and one ancillary mode. The initial matrix corresponds to the unperturbed MMW with $\xi_k=0$ for all tunable elements. Color encodes the optical phase and brightness the field amplitude, as indicated by the complex-field color reference. The optimized transformation reaches a correlation of $99.5\%$ with the target matrix at an optical loss of $L_\eta =0.55$~dB.
    (b) Evolution of the matrix correlation during optimization, approaching the upper bound after approximately $6\times10^3$ iterations.
    (c) Corresponding evolution of the optimization cost function.
    (d) Optimized phase-gradient slopes $\xi_k$ of the $M=10N^2=160$ thermo-optically tunable elements distributed along the MMW. The maximum slope is $|\xi_k|\leq0.26$ and the participation ratio $\epsilon=0.46$.}
    \label{fig:sylvesterN4}
\end{figure}

We next investigate the ability of the thermally programmable MMW to realize a predefined optical transformation.
As an introductory example, we consider normalized $N=4$ Sylvester--Hadamard transformations~\cite{leedumrongwatthanakun2020programmable,mitrouli2014sylvester}.
A Hadamard matrix is composed of elements $\pm1$ and has mutually orthogonal rows and columns.
Sylvester--Hadamard matrices form a particular class that can be constructed recursively as
\begin{equation}
    \mathbf{H}_{2N}
    =\frac{1}{\sqrt{2}}
    \begin{pmatrix}
        \mathbf{H}_N & \mathbf{H}_N\\
        \mathbf{H}_N & -\mathbf{H}_N
    \end{pmatrix}\mathrm{, with\text{ } } \mathbf{H_1}=1.
\end{equation}
The resulting transformation is unitary, such that each input mode is coupled to an equal-amplitude superposition of all four target output modes, with the signs corresponding to relative optical phases of $0$ and $\pi$.
The target transformation is shown in the left panel of Fig.~\ref{fig:sylvesterN4}a, where color encodes the optical phase and brightness the field amplitude.

For the optimization, the first four guided modes define the target modal space, while one additional higher-order guided mode is included as an ancillary mode.
As discussed above, the thermo-optic perturbations induce coupling between the guided modes.
Consequently, any mode within the set of target modes can couple to modes outside that domain during propagation.
If these additional modes are excluded from the model, such coupling is treated solely as leakage from the target modes and cannot be exploited by the optimization.
By accounting for ancillary modes, the optimization can harness this intermediate modal coupling, while the desired input--output transformation remains defined only within the target modes.
This allows optical power to temporarily couple beyond the four target modes during propagation and subsequently couple back.
The number of ancillary modes is chosen according to the coupling range induced by the thermo-optic perturbations.
For the present optimization, the maximum phase-gradient slope is $\xi_{\max}=0.25$, for which coupling occurs predominantly between neighboring modes, as shown in Fig.~\ref{fig:thermalCoupling}.

The programmable MMW contains $M=10N^2=160$ thermo-optically tunable elements, parameterized by the set of phase-gradient strengths
$\boldsymbol{\xi}=\{\xi_1,\ldots,\xi_M\}$.
The initial state corresponds to the unperturbed MMW, for which $\xi_k=0$, for all $k=1,\dots,M$.
The initial transmission matrix is shown in the center panel of Fig.~\ref{fig:sylvesterN4}a.
Although no mode coupling is introduced in this state, the different propagation constants of the guided modes lead to mode-dependent phase accumulation during propagation, resulting in the different phases visible along the main-diagonal of the transmission matrix.

We subsequently optimize all $M$ phase-gradient slopes using a cost function that maximizes the fidelity between the target and obtained output modal vectors.
Additionally, the cost function penalizes propagation loss and power leakage from the target modes. 
A detailed description is provided in the Supplementary Material. 
During optimization, the MMW is probed with different random complex input fields, which are propagated through both the target transformation and the MMW. 
The resulting complex output fields are compared by determining the correlation between the instantaneous and target output modal vectors. 
The cost function additionally considers optical transmission loss and power conservation within the target modes avoiding optical power being coupled to the ancilla mode. 
In our cost function, 
we allow a maximum overall propagation loss of $1$~dB and set the target mode power constraint to $0$~dB with the corresponding weight of $\gamma=0.5$. 

The optimization result is shown in the right panel of Fig.~\ref{fig:sylvesterN4}a and closely reproduces the target Sylvester--Hadamard transformation.
In order to monitor the similarity between  the target and obtained transformation operators, we additionally calculate the full transmission matrix of the MMW, i.e. $\mathbf{TM}_{\mathrm{MMW}}$, every $10^3$ iterations and evaluate its normalized correlation with the target, i.e. $\mathbf{TM}_{\mathrm{target}}$, according to
\begin{equation}
C_{\mathrm{TM}}
=
\frac{
\left|
\sum_{i,j}
\mathbf{TM}_{\mathrm{target,i,j,}}^{\dagger}
\mathbf{TM}_{\mathrm{MMW,i,j,}}
\right|
}{
\left\|\mathbf{TM}_{\mathrm{target}}\right\|_{\mathrm{F}}
\left\|\mathbf{TM}_{\mathrm{MMW}}\right\|_{\mathrm{F}}
},
\label{eq:TM_correlation}
\end{equation}
where $\|\cdot\|_{\mathrm{F}}$ denotes the Frobenius norm.
The evolution of the correlation during optimization is shown in Fig.~\ref{fig:sylvesterN4}b.
The final result reaches a  correlation of $99.5\%$ with the target.
The optimization was performed over $10^4$ iterations and required approximately $55$~s on an NVIDIA RTX 500 Ada Generation Laptop GPU.
The correlation increases rapidly and approaches its upper bound after approximately $6\times10^3$ iterations, demonstrating convergence toward the desired transformation.
The corresponding evolution of the cost function is shown in Fig.~\ref{fig:sylvesterN4}c.

Figure~\ref{fig:sylvesterN4}d shows the distribution of the phase-gradient slopes $\xi_k$ along the MMW after optimization.
Compared with single-spot focusing shown in Fig.~\ref{fig:3}d, the realization of the full $4\times4$ transformation requires stronger perturbations, reflecting the increased complexity of simultaneously controlling whole input--output transformations of the guided modes.
The optimized control distribution has a participation ratio of $\epsilon=0.46$, which is slightly higher than for single-spot focusing, indicating a broader distribution of the control across the available heaters while remaining well below the limit $\epsilon=1$ of uniform participation.
The optimized transformation exhibits an optical loss of the target modes of $L_\eta =0.55$~dB, as defined in the Supplementary Material. 

\section{Scaling to high-dimensional optical transformations}

\begin{figure}[!t]
    \raggedright
    (a)\includegraphics[height=4.8cm]{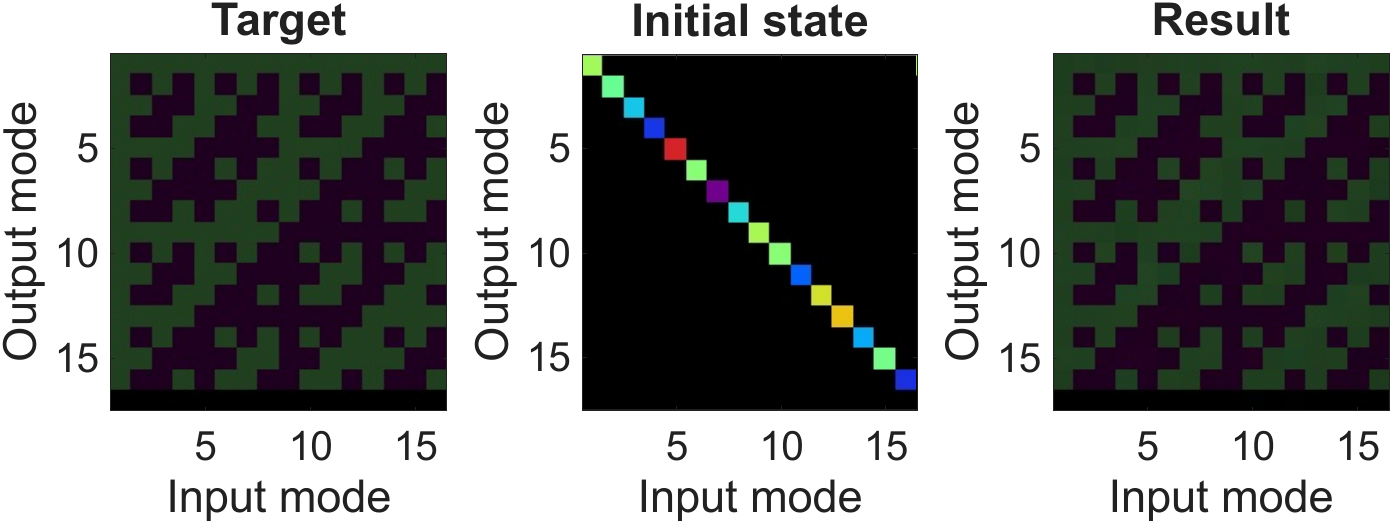}
    \includegraphics[height=4.8cm]{img/complex_colorbar.eps}\\
    (b)\includegraphics[height=4.8cm]{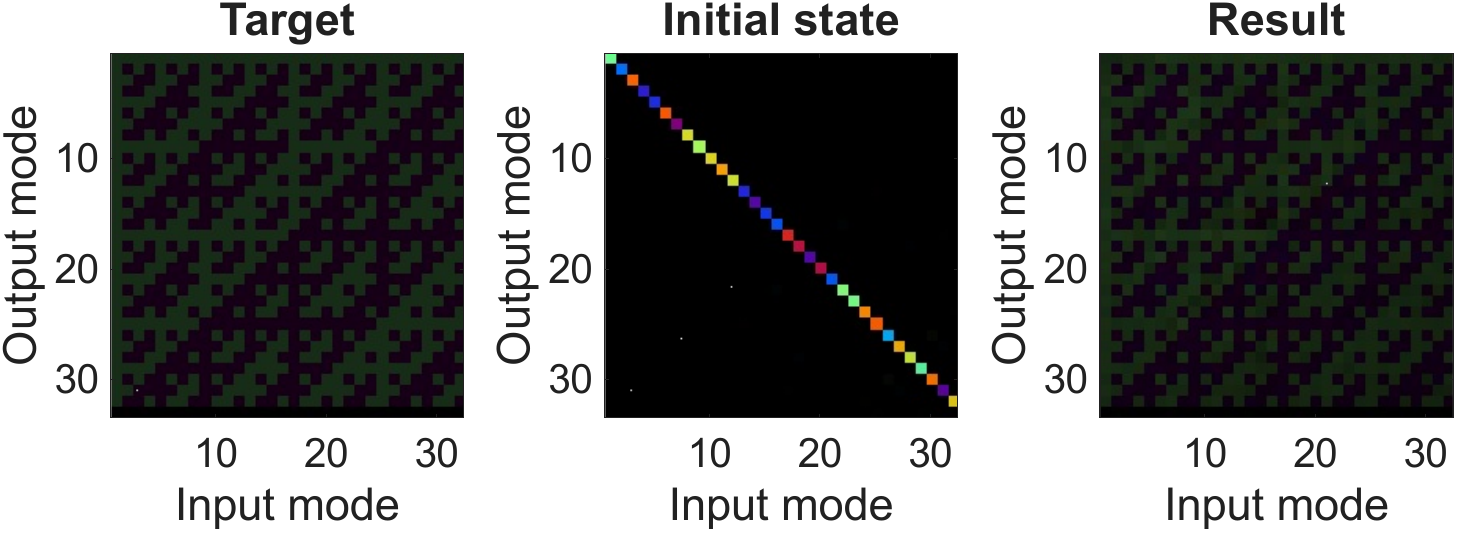}\\  
    \caption{ Scaling to higher-dimensional programmable optical transformations.
    Optimized transmission matrices for (a) $N=16$ and (b) $N=32$ Sylvester--Hadamard transformations using $M=10N^2$ thermo-optically tunable elements and one ancillary mode.
    Color encodes the optical phase and brightness the field amplitude.
    The learning rate for the $N=16$ transformation is $\alpha=2.4\times 10^{-5}$, and for $N=32$ it is $\alpha=8.3\times 10^{-6}$.
    The optimized transformations reach correlations of $99.9\%$ and $99.2\%$, for $N=16$ and $N=32$ modes. For both optimizations the overall optical loss of the target modes is $L_\eta < 0.55$~dB.}
    \label{fig:sylvesterN16_32}
\end{figure}

In the previous section, we demonstrated the realization of a $4\times4$ Sylvester--Hadamard transformation. Next, we investigate the extension of the same architecture to higher-dimensional transformations.
For all considered dimensions, we retain the same scaling of $M=10N^2$ thermo-optically tunable elements and include a single ancillary mode in addition to the $N$ target modes.
As for the $N=4$ case, each optimization is initialized from the corresponding unperturbed MMW, with all phase-gradient strengths set to $\xi_k=0$, for all $k=1,\dots,M$.
Figure~\ref{fig:sylvesterN16_32} shows the optimized transmission matrices for $N=16$ and $N=32$ Sylvester--Hadamard transformations.

For $N=16$, corresponding to $M=2560$ tunable elements, the optimization was performed over $10^5$ iterations.
The resulting transformation reaches a correlation of $99.9\%$ with the target matrix while exhibiting an average optical loss of only $L_\eta< 0.1$~dB.
The largest phase-gradient slope required by the optimized device is $|\xi|_{\max}=0.18$.
The complete optimization required approximately $11.5$~h on the aforementioned GPU.
Increasing the transformation dimension to $N=32$ results in $M=10,240$ independently tunable elements.
After $1.6\times10^6$ optimization iterations, the resulting transformation reaches a correlation of $99.2\%$, while the average optical loss is $L_\eta =0.55$~dB.
The maximum required phase-gradient strength is $|\xi|_{\max}=0.18$.
The optimization required approximately $96$~h.

\section{Reducing the number of tunable elements}

An important consideration in the design of reconfigurable wave systems is the number of independent tuning parameters required to realize a desired functionality. 
Recent work on coherent control of complex scattering systems has shown that this number can, in some cases, be related directly to the number of constraints imposed on the underlying scattering operator and can be substantially smaller than the number of parameters employed in conventional highly parameterized implementations~\cite{alhulaymi2025coherent}.
Here, we investigate how the number of tunable elements in our MMW can be reduced while retaining control over the target optical transformation. 
In the previous section, we demonstrated high-correlation transformations using $M=10N^2$ tunable elements.
We now systematically reduce $M$ for the $N=4$ Sylvester--Hadamard transformation while keeping all other optimization parameters unchanged.
In particular, we initialize the control parameters to zero (i.e. $\xi_k=0$, for all $k=1,\dots,M$), we keep one ancillary guided mode, the same learning rate of $\alpha=2.6\times10^{-4}$. 
We allow a maximum overall propagation loss of $1$~dB and set the target mode power constraint to $0$~dB.
Each configuration is optimized for $2\times10^4$ iterations.

\begin{figure}[!t]
    \raggedright
    (a)\includegraphics[height=4.4cm]{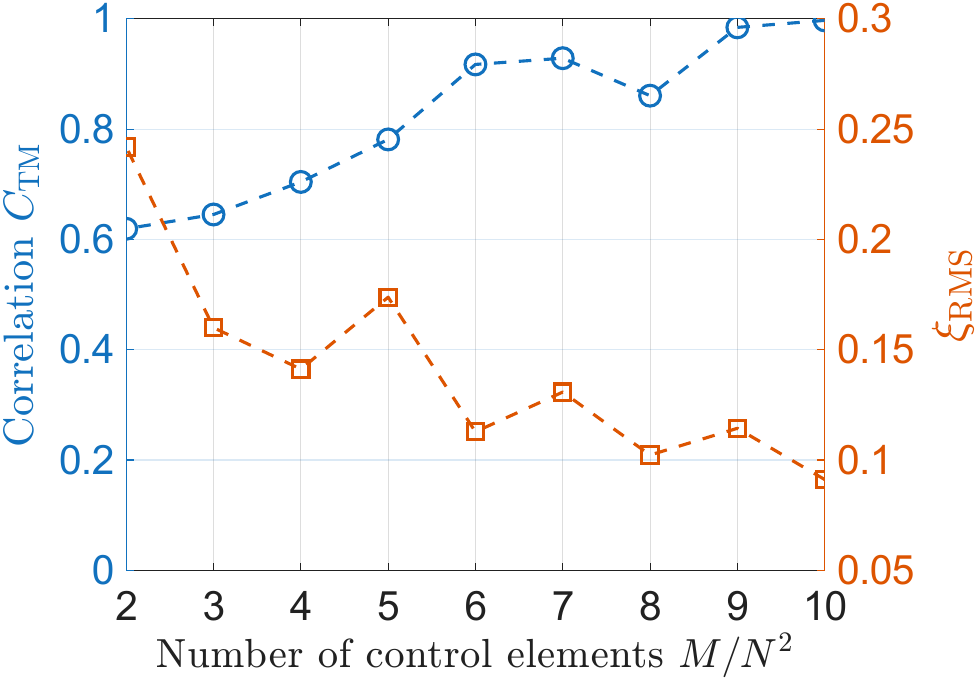} 
    (b)\includegraphics[height=4.4cm]{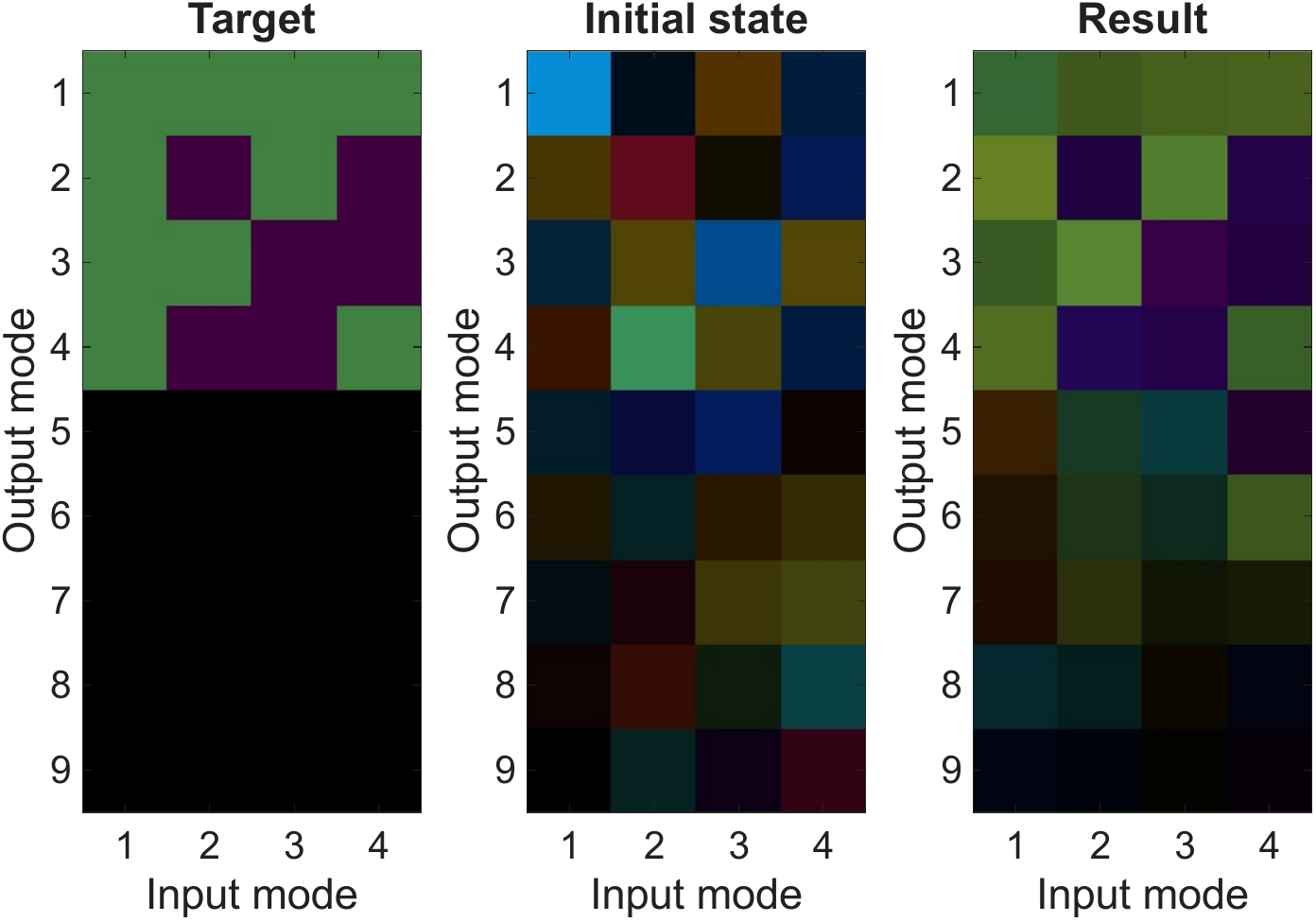} \includegraphics[height=4.2cm]{img/complex_colorbar.eps}  \\
    (c)\raisebox{1.5cm}{\includegraphics[width=0.45\textwidth]{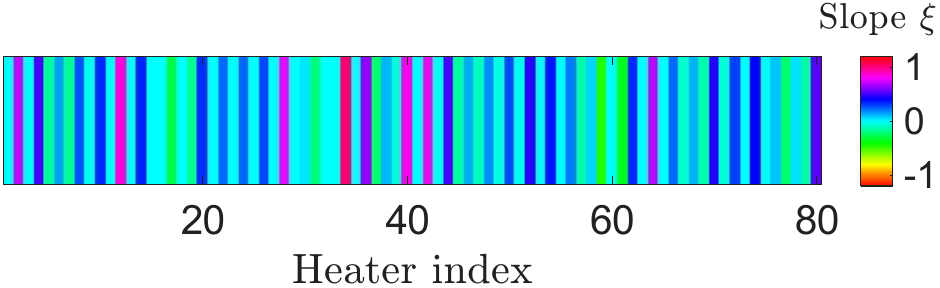}}(d)\includegraphics[width=0.45\textwidth]{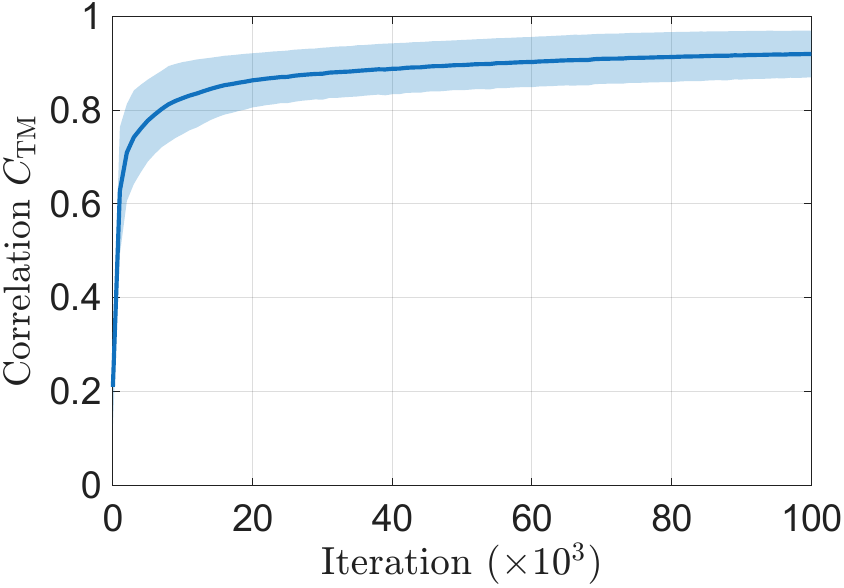} 
    \caption{Reduction of the number of tunable elements.
    (a) Correlation between the target and obtained transformation (left axis) and root-mean-square phase-gradient slope $\xi_{\mathrm{RMS}}$ (right axis) after $2\times10^4$ optimization iterations as a function of the normalized number of tunable elements $M/N^2$ for the $N=4$ Sylvester--Hadamard transformation. All optimizations are initialized with $\xi_k=0$ and use one ancillary guided mode. The learning rate was kept constant at $\alpha=2.6\times10^{-4}$. Reducing the number of tunable elements decreases the achieved correlation while increasing the required phase-gradient slope.
    (b) Target (left) and optimized (right) transmission matrices for one realization with $M=5N^2=80$ tunable elements and five ancillary guided modes. The optimized transformation reaches a correlation of $98.3\%$ with the target matrix and an optical loss of $L_\eta=1.13$~dB.
    (c) Optimized phase-gradient slopes $\xi_k$ corresponding to the
    realization shown in (b). The maximum slope is $|\xi_k|\leq1.030$ and the participation ratio $\epsilon=0.44$.
    (d) Average correlation during optimization over eight independent random
    initializations of the phase-gradient slopes, with the shaded region
    indicating one standard deviation. The eight realizations achieve an
    average correlation of $93.8\% \pm 4.8\%$.
    }
    \label{fig:controlScaling}
\end{figure}

Figure~\ref{fig:controlScaling}a shows the correlation of the optimization result, i.e the effective transformation carried out by the MMW, with the target transformation as a function of the
number of tunable elements.
Reducing $M$ limits the available degrees of freedom and eventually degrades the optimization toward the target transformation.
We find that the correlation progressively drops from $99.5\%$ with $M=10N^2$ to $61.9\%$ with $M=2N^2$ tunable elements.
At the same time, we observe that the optimization increasingly seeks stronger individual perturbations when reducing the number of tunable elements.
This is quantified in Fig.~\ref{fig:controlScaling}a by the root-mean-square of the optimized gradient slopes $\xi_{\mathrm{RMS}}=\sqrt{M^{-1} \sum_{k=1}^M \lvert \xi_k \rvert^2}$, which increase by reduction of $M$.
As discussed in Fig.~\ref{fig:thermalCoupling}, increasing the perturbation strength broadens the induced modal coupling.
Consequently, stronger perturbations increasingly couple optical power to higher-order guided modes outside the $N$ target modes, as well as to radiation modes that are no longer confined by the MMW.

These observations suggest that the additional modes accessed by the stronger perturbations may themselves be incorporated into the optimization rather than treated solely as unwanted leakage.
We therefore next investigate whether high-fidelity transformations can still be realized with a substantially reduced number of tunable elements.
Specifically, we consider only $M=5N^2=80$ tunable elements.
To account for the broader modal coupling associated with the stronger perturbations, we increase the number of ancillary guided modes from one to five.
We additionally set the maximum overall propagation loss to $-200$~dB and set the target mode power constraint to $-5$~dB.
We further find that with a reduced number of tunable elements the overall convergence becomes increasingly sensitive to the initial control configuration.
While the aforementioned transformations were initialized from the unperturbed MMW with $\xi_k=0$, for all $k=1,\dots,M$, we now initialize the phase-gradient slopes $\boldsymbol{\xi}$ randomly around a relatively small slope of $0.3$. 
In order to evaluate the robustness with respect to the initialization, we perform eight independent optimizations with randomly chosen initial configurations. Figure~\ref{fig:controlScaling}d shows the average correlation and corresponding standard-deviation envelope over the first $10^5$ optimization iterations. Across the eight realizations, we obtain an average correlation of $93.8\% \pm 4.8\%$. 
Individual realizations were continued beyond $10^5$ iterations where required to reach convergence. The realization shown in Fig.~\ref{fig:controlScaling}b reaches a correlation of $98.3\%$ with the target transformation.
The corresponding optimized gradient slopes are shown in Fig.~\ref{fig:controlScaling}c.
The maximum phase-gradient slope increases substantially to $\xi_{\max}=1.03$, confirming that reducing the number of tunable elements requires locally stronger perturbations. 
At the same time, the participation ratio remains relatively low at $\epsilon=0.43$, indicating that these strong perturbations occur locally, rather than being distributed uniformly across all tunable elements. 
This limited average participation is consistent with the moderate optical loss of $L_\eta=1.13$~dB.

In order to verify that the reduced-control configuration is not specific to a single target transformation, we additionally optimize a rotated $4\times4$ transformation using the same number of tunable elements. 
This transformation also reaches a correlation of $98.27\%$, as shown in Fig.~\ref{fig:rotatedTransformation}.

\section{Discussion and conclusion}

We have shown that thermally perturbed integrated multimode waveguides can be harnessed to precisely control the propagation of optical fields while maintaining low optical loss within the considered model. 
We developed a physical model that describes the interaction between distributed thermo-optic perturbations and the multimode propagation. 
Consequently, the search for the required control settings constitutes a multivariable optimization problem.
Here, we employed gradient-based optimization using a forward model.
We note that alternative approaches such as genetic algorithms or neural networks could similarly be applied~\cite{feng2019multi}.

We first demonstrated the concept for integrated wavefront shaping by focusing light at the output facet of the multimode waveguide.
For $N=20$ excited modes, the optimization reaches the theoretical focusing limit of $\approx 84\%$ of optical output power concentrated within the focal area, by using 80 tunable elements.
Such integrated wavefront shaping could be particularly attractive for LiDAR~\cite{park2021all} and free-space optical communication systems~\cite{xie2025terabits}, where our programmable multimode waveguide could, for example, steer and focus the transmitted beam toward a receiver to maximize the coupled optical power. 
This is particularly relevant for platforms that require compact and lightweight components such as satellite laser terminals~\cite{wang2024free}.

Conventional wavefront-shaping devices, such as spatial light modulators, typically have a fixed number and arrangement of pixels.
In contrast, in our perturbed multimode waveguide the number and spatial distribution of the tunable elements can be adapted to the requirements of the desired optical functionality.

Beyond intensity shaping, we demonstrated programmable optical transformations using the same perturbed multimode waveguide.
Sylvester--Hadamard transformations with dimensions up to $N=32$ were realized with correlations of up to $99.9\%$ and low optical loss well below $1$~dB. 
This provides an alternative to universal linear optical processors based on meshes of single-mode Mach--Zehnder interferometers~\cite{reck1994experimental,clements2016optimal,carolan2015universal,Taballione2019reconfigurable}. 
Such architectures require a number of tunable elements scaling as $\mathcal{O}(N^2)$ and rely on increasingly large networks of interferometric building blocks with accurately controlled splitting ratios and relative phases as the transformation dimension increases.
Our approach does not eliminate the need for the quadratic control element scaling, but considerably simplifies the underlying photonic architecture.
All spatial modes co-propagate within the same waveguide, and the only distributed tunable components are the thermo-optic heaters.
Importantly, these tunable elements do not need to provide identical perturbations, since variations in their individual responses can be accounted for during optimization.

The required number of thermo-optic  perturbations $M$ represents a fundamental design parameter of the programmable MMW. 
More generally, identifying the number of independent tuning parameters required for a prescribed wave transformation is an important problem in reconfigurable wave systems.
Recent theoretical work has related such minimum parameter counts to the number of independent constraints imposed on the scattering operator through codimension arguments~\cite{alhulaymi2025coherent,guo2023singular}.
These works emphasize the importance of matching the dimensionality of the available control space to that of the desired optical functionality.
This distinction is also reflected in our results.
For single-spot focusing using $N=20$ modes, $M=N$ and $M=2N$ tunable elements reach power ratios of approximately $56\%$ and $78\%$, respectively, whereas $M=4N$ is sufficient to reach the theoretical focusing limit of approximately $84\%$.
For intensity shaping using $N$ modes, an orthogonal control basis of the same dimensionality would in principle provide sufficient independent degrees of freedom to achieve the optimum focusing performance.
The thermo-optic phase gradients employed here do not form such an orthogonal basis, resulting in partially redundant control and therefore requiring a larger number of tunable elements.

For a general complex $N\times N$ linear transformation, the number of independent parameters itself scales as $\mathcal{O}(N^2)$, such that a corresponding scaling of the available control degrees of freedom is generally expected for arbitrary transformations.
The number of tunable elements required to provide these degrees of freedom depends on how effectively and independently each element controls the optical transformation.
For our multimode waveguide, using $M=10N^2$ tunable elements allows the realized transformations to approach unit correlation with the target.
As $M$ is reduced, the optimization compensates for the smaller number of available controls through stronger individual perturbations.
The resulting stronger modal coupling extends over a broader range of modes, including guided modes outside the target modes, which we explicitly include as ancillary modes in the optimization.
Stronger perturbations can additionally increase coupling to radiation modes.
By including additional ancillary modes and relaxing the loss constraint, we obtain an average correlation of $93.8\%\pm4.8\%$ over eight random initializations using $M=5N^2$ tunable elements, with the best-performing realization reaching $98.3\%$ correlation with the target.
The number of tunable elements may be reduced further through perturbation profiles that provide more effective and independent control of the modal field and through optimized longitudinal placement of the heaters, for example by placing heaters directly above the multimode waveguide~\cite{cheng2024multimodal}.

An important consideration for scaling the proposed architecture is the total optical path length, which is directly associated with the overall optical loss.
For the largest transformation considered here, $N=32$, the present design uses $M=10N^2=10,240$ tunable elements.
Assuming a heater length of $2$~mm and a heater separation of $50$~\textmu m, a direct implementation would correspond to a total optical path length of approximately $21$~m.
Although such a length is substantial for an integrated device, meter-scale optical path lengths have been demonstrated~\cite{belt2017ultra} in integrated waveguides.
Ultra-low-loss $\mathrm{Si}_3\mathrm{N}_4$ has demonstrated propagation losses on the order of $0.1$~dB/m~\cite{roeloffzen2018low}, which would correspond to approximately $2.1$~dB of propagation loss over the maximum propagation length considered in this work.
Importantly, the dimensions used in the present model have not been optimized for device footprint.
The required interaction length of each tunable element is directly related to the achievable thermo-optic phase perturbation and could therefore be substantially reduced by employing materials with larger thermo-optic coefficients.
Polymer-based waveguides, for example, can exhibit thermo-optic coefficients on the order of $10^{-4}$~K$^{-1}$~\cite{zhang2006thermo}, substantially larger than those of conventional ${\mathrm{Si}_3\mathrm{N}_4}$, and could therefore enable potentially shorter interaction lengths, approximately scaling inversely with the thermo-optic coefficient.
Furthermore, the present architecture employs a single transverse perturbation profile at each longitudinal position.
Introducing multiple independently controlled heaters across the transverse direction~\cite{cheng2024multimodal} would provide several control degrees of freedom within the same longitudinal section.
This could noticeably reduce the number of sequential tuning sections required for a given transformation and thereby shorten the overall optical path length.

Our proposed architecture shows spatial shaping of a beam that propagates through a multimode waveguide. This concept could be extended by for instance coupling ring resonators upstream/downstream of the multimode waveguide~\cite{bogaerts2012silicon}.
Combining spectral and spatial control would enable spatio-spectral and, for coherent broadband fields, spatio-temporal shaping within an integrated platform~\cite{sun2018four}. 

The demonstrated low-loss optical transformations may be relevant for quantum photonic applications.
Passive linear transformations of spatial modes can equally be applied to single-photon and multi-photon states.
Related programmable transformations have previously been demonstrated using externally controlled large-scale multimode fibers~\cite{defienne2016two, wolterink2016programmable,leedumrongwatthanakun2020programmable}. 
A fully integrated implementation could therefore enable high-dimensional quantum transformations robustly and, in combination with
photon-pair sources, the generation and manipulation of spatially
entangled states~\cite{valencia2020unscrambling}.

The present results are numerical and several aspects will require experimental validation. 
In particular, a fabricated device will exhibit wavelength and polarization-dependent mode fields, fabrication imperfections and cross-talk between the thermo-optically tunable elements, that are only approximately represented by the present model. 
Although these effects could be incorporated into more complete electromagnetic and thermal models, our differentiable model does not need to perfectly reproduce the fabricated device, as it could provide an initial solution that is subsequently refined using measured optical feedback.

\subsection*{Funding}
This work was supported by Horizon Europe Project Qu-PIC under Grant 101135845. IMV is funded by NWO Vici project number 21646 under the grant https://doi.org/10.61686/YNCIU74074.

\subsection*{Acknowledgement}
We thank Chun-Wei Chen at University of Bath for stimulating discussions.

\subsection*{Disclosure}
The authors declare no conflicts of interest.

\subsection*{Data availability}
The source code and data underlying the presented results will be made publicly available upon publication.

\subsection*{Supplemental document}
See Supplement 1 for supporting content.

\clearpage
\renewcommand{\appendixname}{Supplementary Material}
\renewcommand{\appendixpagename}{Supplementary Material}
\renewcommand{\appendixtocname}{Supplementary Material}

\begin{appendices}
\renewcommand{\thesection}{\arabic{section}}
\setcounter{figure}{0}
\renewcommand{\thefigure}{S\arabic{figure}}

\renewcommand{\theequation}{S\arabic{equation}}
\setcounter{equation}{0}

\section*{Modal properties of the unperturbed multimode waveguide}
\begin{figure}[h]
    \centering
    (a)\includegraphics[height=4cm]{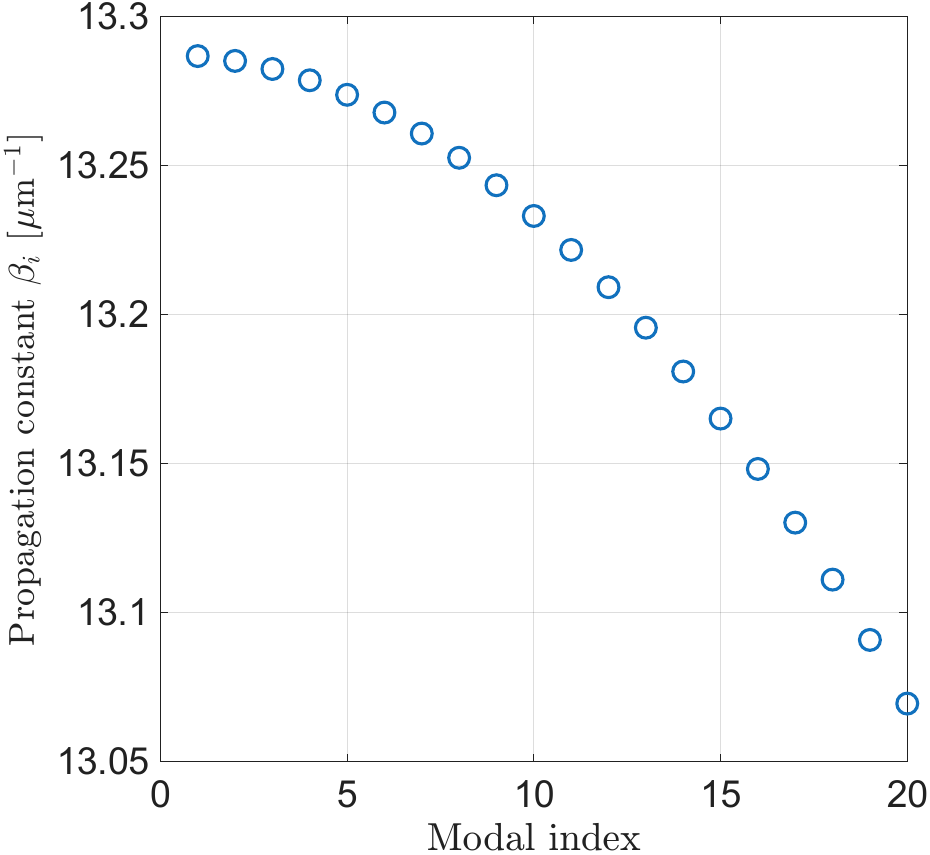}
    (b)\includegraphics[height=4cm]{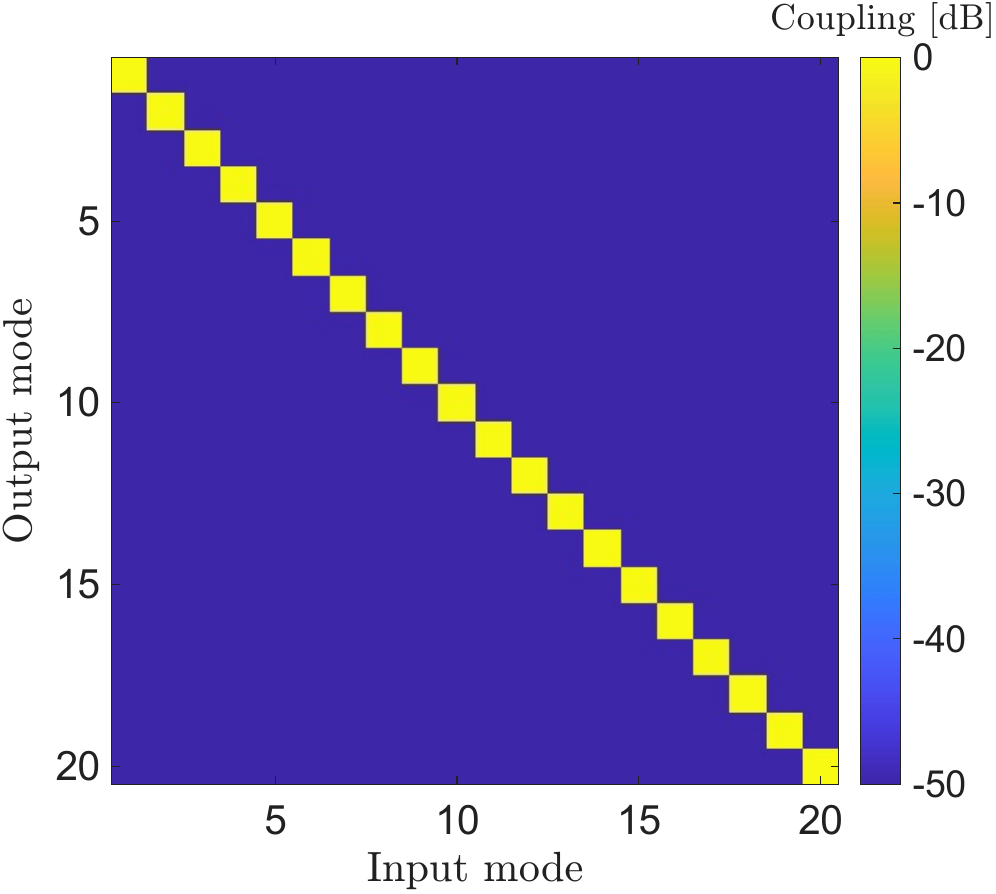}
    \caption{{Modal properties of the unperturbed MMW.}
    (a) Propagation constants $\beta_i$ of the first 20 guided spatial modes.
    (b) Normalized modal overlap matrix $C_{ij}$. 
    The strong main-diagonal elements confirm the orthogonality of the calculated mode set.}
    \label{fig:modeSet}
\end{figure}

The mode profiles and propagation constants of the unperturbed multimode waveguide~(MMW) were calculated using the eigenmode solver of Lumerical MODE, as described in the Methods section in the main part. 
Figure~\ref{fig:modeSet}a shows the propagation constants $\beta_i$ of the first 20 guided spatial modes. 
To verify the orthogonality of the calculated mode set, we evaluate the normalized modal overlap matrix
\begin{equation}
C_{ij}
=
\frac{
\left|
\iint E_i(x,y) E_j^*(x,y)\,\mathrm{d}x\,\mathrm{d}y
\right|^2
}{
\left(\iint |E_i(x,y)|^2\,\mathrm{d}x\,\mathrm{d}y\right)
\left(\iint |E_j(x,y)|^2\,\mathrm{d}x\,\mathrm{d}y\right)
},
\end{equation}
where $E_i(x,y)$ and $E_j(x,y)$ denote the transverse electric fields of modes $i$ and $j$, respectively.
As shown in Fig.~\ref{fig:modeSet}b, the overlap matrix is diagonal, confirming the mutual orthogonality of the calculated guided modes.

\section*{Waveguide geometry and mode calculations}

All numerical investigations were performed for silicon nitride ($\mathrm{Si}_3\mathrm{N}_4$) waveguides embedded in a $\mathrm{SiO}_2$ cladding. 
The cladding dimensions were chosen sufficiently large to ensure negligible optical leakage loss. 
The waveguide core width and height were fixed at $26$~\textmu m and $400~\mathrm{nm}$, respectively. 
The same waveguide geometry and corresponding set of guided modes were used throughout the different numerical investigations. 
To study systems with different numbers of spatial modes, only a subset of the available guided modes was excited and considered in the respective simulation. 

The mode profiles and corresponding propagation constants $\beta_m$ were calculated using the eigenmode solver of Lumerical MODE. 
The resulting guided modes form an orthogonal modal basis, as verified by the modal overlap matrix shown in Fig.~\ref{fig:modeSet}b of the Supplementary Material. 
The corresponding propagation constants of the first $20$ guided modes are shown in Fig.~\ref{fig:modeSet}a, illustrating the mode-dependent propagation constants that give rise to modal dispersion during propagation through the MMW.
Figure~\ref{fig:1}b shows the normalized electric-field distributions $\mathrm{Re}\{E(x,y)\}$ of the first three guided modes of the unperturbed MMW.

\section*{Deatils on the thermo-optic simulations}

Here, we provide additional details on the finite-element thermal simulations
used to determine the temperature distributions shown in Fig.~\ref{fig:2} of
the main text.
Thermo-optic control was implemented using platinum~(Pt) heaters positioned adjacent to the MMW core. 
Unless stated otherwise, the heaters were placed at a distance of $3$~\textmu m from the edge of the core.
Despite the close proximity of the heater to the MMW core, the guided modes experience negligible optical absorption. 
This is due to the strong modal confinement in the MMW, which keeps the overlap between the optical modes and the metallic heater relatively small. 

The resulting temperature distributions were calculated using finite-element simulations in COMSOL Multiphysics.
The heater and MMW were embedded in a $\mathrm{SiO}_2$ environment. We use $\mathrm{Si}_3\mathrm{N}_4$ with thermo-optic coefficient of $\frac{\partial n_{\mathrm{Si}_3\mathrm{N}_4}}{\partial T}=2.45\cdot 10^{-5}~\mathrm{K}^{-1}$.
The upper cladding layer was exposed to a $5$-\textmu m-thick layer of air, while the bottom boundary was maintained at room temperature, i.e. $293.3~\mathrm{K}$, reflecting a Peltier element. 
The horizontal simulation boundaries were modeled as ideal black-body radiators to account for thermal radiation losses.
Applying a voltage $U_o$ to the heater generated localized Joule heating, resulting in a spatially varying temperature profile across the waveguide cross section. Figure~\ref{fig:2}a shows a representative temperature distribution applying $2U_o$, while Fig.~\ref{fig:2}b presents the corresponding temperature profiles $\Delta T$ across the waveguide core for three different heater voltages ($U_0$, $2U_0$, and $3U_0$).
The representative heater geometry considered in this work is described below. 

\section*{Heater design}

\begin{figure}[h]
    \raggedright
    (a)\includegraphics[height=4.5cm,trim=4cm 0.5cm 3cm 1cm,
    clip]{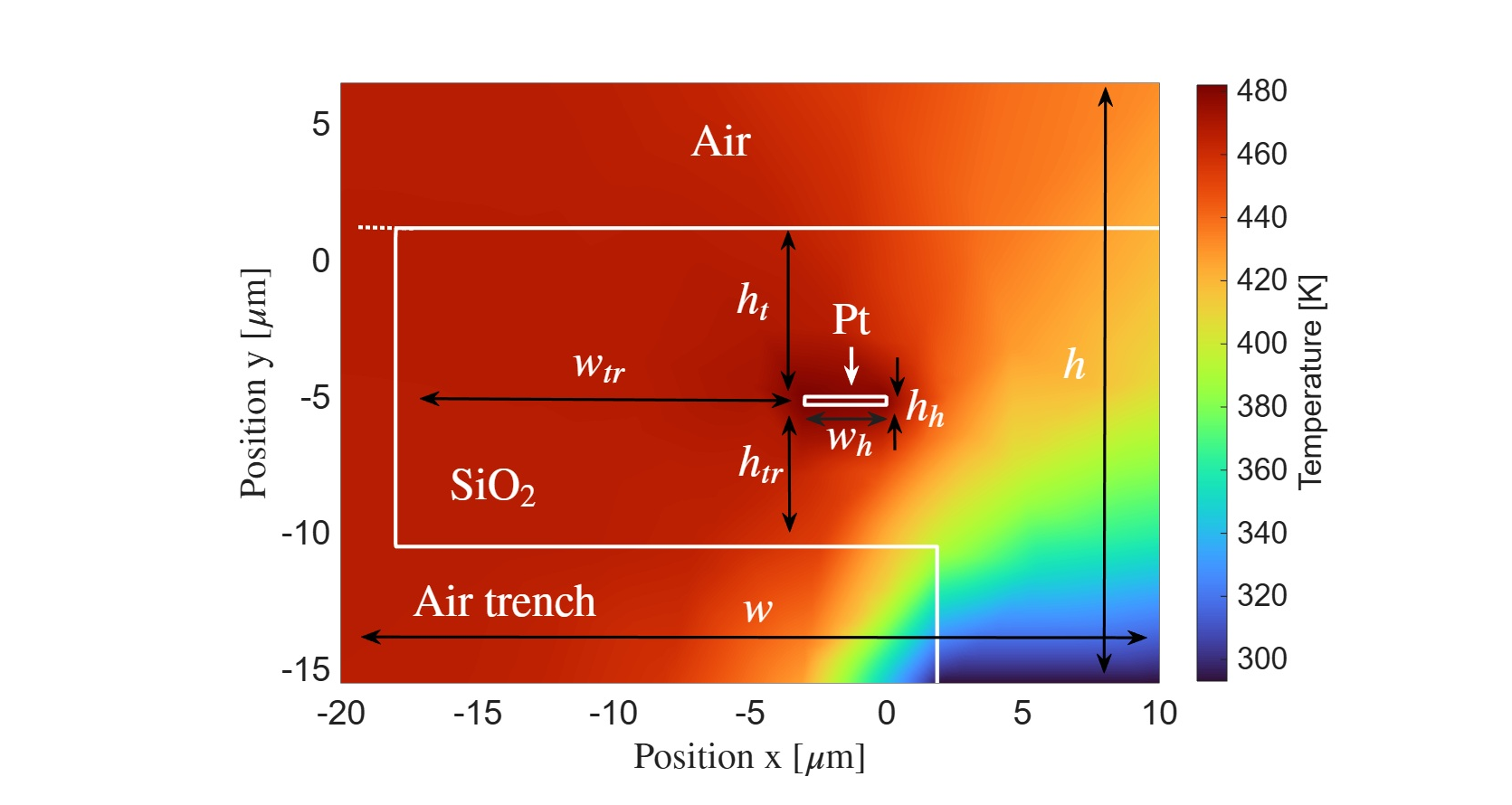}(b)\includegraphics[width=3.5cm]{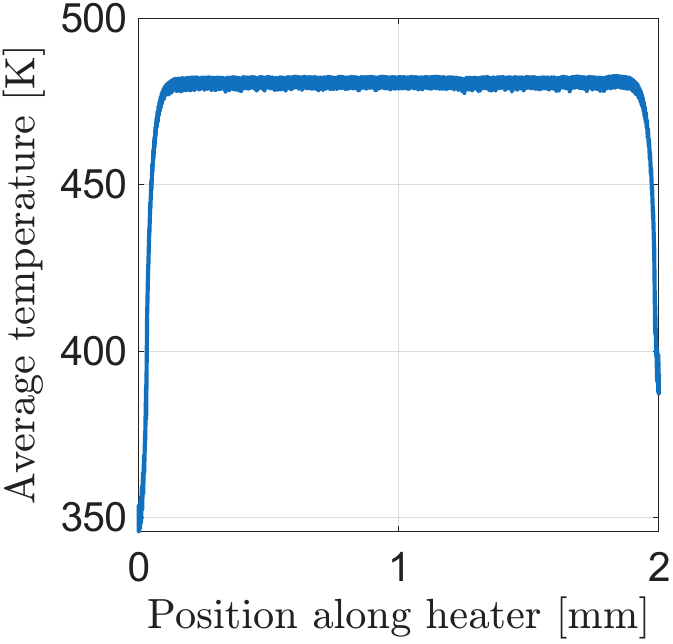}
    (c)\includegraphics[width=3.5cm]{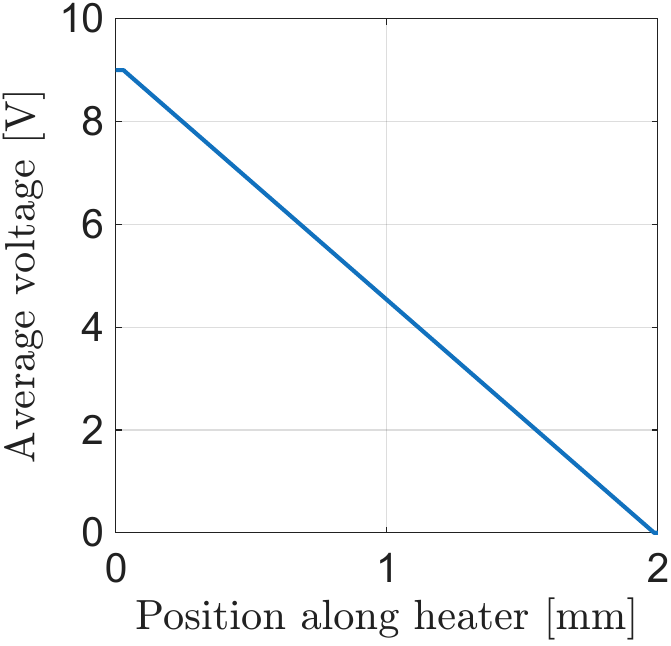} 
    \caption{Full platinum (Pt) heater design and electro-thermal simulation. (a) Cross section of the simulated heater geometry. A $3$-\textmu m-wide Pt heater is embedded in $\mathrm{SiO}_2$ adjacent to the $\mathrm{Si}_3\mathrm{N}_4$ MMW (not shown), with air above the cladding and trenches adjacent to and beneath the heater providing thermal isolation. The dashed line marks the interface between the top surface of the PIC and ambient air. (b) Cross-section-averaged temperature along the $2$-mm-long Pt heater for an applied voltage of $9$~V, showing nearly uniform heating of the active section to approximately $480$~K. (c) Corresponding cross-section-averaged electric potential along the heater, showing the continuous voltage drop from the $9$-V input terminal to ground.}    
\label{fig:heater3D}
\end{figure}

In order to estimate a realistic heater geometry and power consumption, we simulate a full platinum heater design capable of producing the thermo-optic phase shift required for programmable control. Using $\frac{dn_{\mathrm{Si}_3\mathrm{N}_4}}{d T}=2.45\cdot10^{-5}~\mathrm{K}^{-1}$, a $2\pi$ phase shift at $\lambda=785$~nm wavelength requires a temperature difference of approximately $\Delta T=32$~K over a $1$-mm interaction length, according to Eq.~(\ref{eq:phase-shift}) in the main text. For our simulation environment, we therefore consider a $2$-mm-long heater as a technologically conservative geometry that provides sufficient phase-shift margin.

Figure~\ref{fig:heater3D}a shows the simulated heater cross section. The heater is implemented in a thermally isolated suspended geometry, which has previously been shown to enable highly power-efficient thermo-optic phase shifting~\cite{yong2022power}. Specifically, the platinum heater is embedded in a $\mathrm{SiO}_2$ cladding, while air is present above the cladding and in trenches adjacent to and beneath the heater to suppress thermal leakage. This thermal isolation improves power efficiency by confining the generated heat to the active region.

Applying $9$~V heats the narrow heater section to approximately $480$~K, corresponding to a total dissipated electrical power of $331$~mW. Figure~\ref{fig:heater3D}b shows the cross-section-averaged temperature distribution along the $2$-mm-long heater, demonstrating nearly uniform heating throughout the active section. For operation at $\lambda=785$~nm, the resulting thermal gradient provides approximately $2.5\times2\pi$ phase modulation. 
Assuming linear thermal scaling, this corresponds to an estimated $P_{2\pi}\approx132$~mW and $P_{\pi}\approx66$~mW.
Figure~\ref{fig:heater3D}c shows the corresponding electric potential along the heater, exhibiting the continuous voltage drop from $9$~V to ground expected for resistive Joule heating.

\section*{Nonlinear response to individual thermo-optic perturbations}

\begin{figure}[h]
    \centering
    \includegraphics[height=5cm]{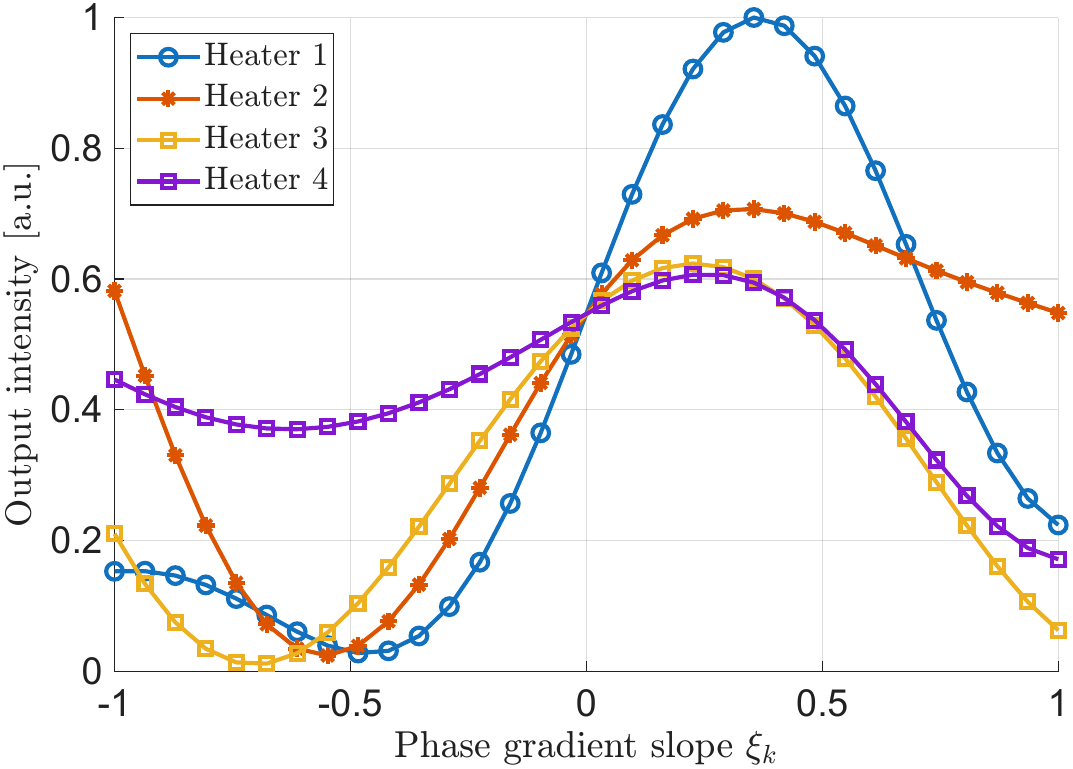}

    \caption{Output response of the MMW to individual thermo-optic perturbations. Starting from the unperturbed MMW with $\xi_k=0$ for all control elements, the phase-gradient strength of each of the first five tunable elements is varied individually while all remaining elements are kept at $\xi=0$. The normalized intensity within the central target region at the output facet is shown as a function of $\xi_k$. For small perturbations around $\xi_k=0$, the response is approximately linear, whereas stronger perturbations produce increasingly pronounced nonlinear behavior. The different responses arise because the tunable elements act at different longitudinal positions and therefore on different modal superpositions.}
\label{fig:single_heater_response}
\end{figure}

To illustrate the nonlinear dependence of the MMW output field on the thermo-optic control parameters, we independently vary the strength of individual perturbations while monitoring the optical intensity within a fixed target region at the MMW output facet.
Initially, all control parameters are set to zero,
$\xi_k=0$ for $k=1,\ldots,M$, corresponding to the unperturbed MMW.
The dimensionless parameter $\xi_k$ parametrizes the strength of the thermo-optic perturbation induced by the $k$th tunable element, as described in Eq.~(\ref{eq:phase_gradient}) in the main text.
The phase-gradient strength $\xi_k$ of one tunable element is then varied while all remaining elements are kept at $\xi=0$.
For each value of $\xi_k$, the optical field is propagated to the MMW output and the intensity within the central target region used for single-spot focusing is evaluated.
This procedure is repeated independently for the first four tunable elements.
Figure~\ref{fig:single_heater_response} shows the resulting normalized output intensity as a function of $\xi_k$.
For small variations of $\xi_k$ around the unperturbed state, the output intensity follows an approximately linear dependence on the perturbation strength.
As $|\xi_k|$ increases, corresponding to increasingly strong thermo-optic perturbations, the response progressively deviates from this local linear behavior and develops pronounced nonlinear features.
This illustrates that, while the response to sufficiently weak perturbations can locally be approximated as linear, the dependence of the propagated field on the thermo-optic control parameters becomes increasingly nonlinear for stronger perturbations.
The response furthermore differs between the individual tunable elements because they are located at different longitudinal positions and therefore act on different modal superpositions of the propagating field.

\section*{List of different cost functions for gradient-descent optimization}
\subsection*{Cost function for single-spot focusing}
In Fig. 3 we show wavefront shaping using our programmable MMW. Our goal is to concentrate the output optical power into a predefined focal region at the facet of the waveguide, as shown in seminal papers on wavefront shaping~\cite{vellekoop2007focusing,vellekoop2008phase,popoff2010}. In order to set up our cost function, we specify a target region $\Omega_{\mathrm{target}}$ around the desired focal position. The size of the focal region is determined from the set of modes propagating in our MMW. We first select a target pixel as the center of the desired focus. This target field is then projected onto the modal basis of the waveguide (see exemplary projection in Eq.~[\ref{eq:modal_decomp}]), yielding the modal coefficients required to create the target focus. The corresponding intensity profile is reconstructed from these modal coefficients and used to specify $\Omega_{\mathrm{target}}$ via the full width at half maximum~(FWHM). The target field is subsequently truncated outside this region and projected onto the modal basis again. After approximately three iterations, the focal region converges to a tightly confined spot. This procedure establishes the theoretical upper bound on the focusing performance achievable with the available modal basis. The cost function is then defined through the power ratio (PR)~\cite{gomes2022near}, as introduced in Eq.~(\ref{eq:pr}) of the main part, which corresponds to the fraction of the total optical power contained within the target region. 
Maximizing the power ratio therefore increases the concentration of optical power at the desired output location. 
Since thermo-optic perturbations may couple optical power into radiation modes, we include the normalized output power $\frac{P_{\mathrm{out}}}{P_{\mathrm{in}}}$ as a second optimization objective to penalize propagation loss.
The cost function for single-spot focusing is therefore defined as:
\begin{equation}
\mathcal{L}=1-\mathrm{PR}-\gamma \frac{P_{\mathrm{out}}}{P_{\mathrm{in}}},
\label{eq:cost_single_spot_focusing}
\end{equation}
where $\gamma$ controls the relative weight assigned to optical transmission.

\subsection*{Cost function for multi-spot focusing}

In addition to single-spot focusing, our programmable MMW can generate multiple target foci on the output facet, simultaneously. To this end, we define a set of non-overlapping target focal regions $\Omega_1,\Omega_2,\ldots,\Omega_P$. The size of each focal region is determined using the same procedure as described for single-spot focusing. 
For each target region, we calculate an individual $\mathrm{PR}$ (Eq.~(\ref{eq:pr})), excluding the power within all focal regions from the total power.
Our objective for this task is twofold: maximize the optical power delivered to the focal regions while simultaneously ensuring a uniform power distribution among all foci. We therefore define the average power ratio

\begin{equation}
\overline{\mathrm{PR}}
=
\frac{1}{P}
\sum_{j=1}^{P}
\mathrm{PR}_j ,
\end{equation}

and the corresponding standard deviation

\begin{equation}
\sigma_{\mathrm{PR}}
=
\sqrt{\frac{1}{P-1}
\sum_{j=1}^{P}
\left( \mathrm{PR}_j-\overline{\mathrm{PR}} \right)^2
}.
\end{equation}

The optimization cost function is then chosen as 

\begin{equation}
\mathcal{L} = 1 - \overline{\mathrm{PR}} + \gamma \sigma_{\mathrm{PR}},
\end{equation}

with the empirically determined weighting factor of $\gamma=0.01$.

\subsection*{Cost function for customized transformation operators}

Beyond wavefront shaping, we investigate whether the programmable MMW can realize arbitrary optical transformations.
We define a $N\times N$ target transformation $\mathbf{TM}_{\mathrm{target}}$, which is
chosen to be unitary,
\begin{equation}
    \mathbf{TM}_{\mathrm{target}}
    \mathbf{TM}_{\mathrm{target}}^\dagger
    = \mathbf{I},
\end{equation}
where $\mathbf{I}$ denotes the identity matrix and $\dagger$ the Hermitian
transpose.
Our objective is to determine the set of phase-gradient slopes $\boldsymbol{\xi}=(\xi_1,\ldots,\xi_M)$ such that the effective transformation of the MMW, $\mathbf{TM}_{\mathrm{MMW}}(\boldsymbol{\xi})$, reproduces the desired target
transformation. 
We sample the instantaneous transformation by using random superpositions
of the target input modes. At each optimization iteration $j$, a random-phase
input vector is generated according to
\begin{equation}
    \mathbf{E}_{\mathrm{in},j} =
    \left(
    e^{i\phi_{1,j}},\ldots,e^{i\phi_{N,j}}
    \right)^{\mathrm{T}},
\end{equation}
where the phases $\phi_{n,k}$ are independently and uniformly drawn from $[0,2\pi]$. 
As such, all target input modes have equal amplitude and statistically independent phases. 
If ancillary modes are included in the simulation, their input
amplitudes are set to zero. 
The corresponding desired output modal vector is given by
\begin{equation}
    \hat{\mathbf{y}}_j
    =
    \mathbf{TM}_{\mathrm{target}}\mathbf{E}_{\mathrm{in},j}.
\end{equation}
The same input vector is propagated through the differentiable forward model of the programmable MMW, yielding
\begin{equation}
    \mathbf{b}_j
    =
    \mathbf{TM}_{\mathrm{MMW}}(\boldsymbol{\xi})\mathbf{E}_{\mathrm{in},j}.
\end{equation}
The agreement between the desired and obtained output vectors is
quantified by their normalized squared correlation, which we denote as the instantaneous fidelity,
\begin{equation}
    F_j =
    \frac{
    \left|
    \hat{\mathbf{y}}_j^\dagger \mathbf{b}_j
    \right|^2
    }{
    \left\|\hat{\mathbf{y}}_j\right\|_2^2
    \left\|\mathbf{b}_j\right\|_2^2
    }.
\end{equation}
Consequently, $0\leq F_j\leq1$, with $F_j=1$ corresponding to identical target and obtained output modal vectors.
In addition to maximizing the fidelity, two power constraints are incorporated into the optimization. 
The total guided-power transmission is defined as
\begin{equation}
    \eta_{\mathrm{out},j}
    =
    \frac{P_{\mathrm{out}}}{P_{\mathrm{in}}},
\end{equation}
and accounts for optical power that remains within all guided modes, including the ancilla modes. 
The fraction of this guided output power that remains within the $N$ target modes is given by
\begin{equation}
    \eta_{\mathrm{target},j}
    =
    \frac{
    \left\|\mathbf{b}_{\mathbf{T},j}\right\|_2^2
    }{
    P_{\mathrm{out}}
    }.
\end{equation}
The instantaneous cost function evaluated at iteration $j$ is
\begin{equation}
    \mathcal{L}_k
    =
    \left(1-F_j\right)
    +
    \max\!\left(
        0,\eta_{\mathrm{out,min}}-\eta_{\mathrm{out},j}
    \right)
    +
    \gamma_t
    \max\!\left(
        0,\eta_{\mathrm{target,min}}-\eta_{\mathrm{target},j}
    \right),
\end{equation}
where $\eta_{\mathrm{out,min}}$ specifies the minimum accepted power transmission, $\eta_{\mathrm{target,min}}$ the minimum fraction of guided output power retained within the target modes, and $\gamma_t$ controls the relative weight of the latter contribution. 
The transmission thresholds are defined as loss limits in decibels according to
$\eta_{\min}=10^{L_{\max}/10}$, where $L_{\max}\leq0$ denotes the specified loss limit in dB. 

At each iteration, a new statistically independent random-phase input vector is generated and the gradient of $\mathcal{L}_j$ with respect to all phase-gradient slopes $\boldsymbol{\xi}$ is calculated. 
The control parameters are updated using the Adam optimizer. 

In order to monitor the convergence of the complete optical transformation, the full transmission matrix is calculated at regular intervals during the optimization. 
The correlation values reported in the convergence plots as in Fig.~\ref{fig:sylvesterN4} and Fig.\ref{fig:sylvesterN16_32} quantify the normalized correlation between the target transmission operator and the transmission operator
realized by the MMW.

\section*{Initial estimate of a stable learning rate}
The maximum stable learning rate was determined as a function of the number of control elements $M$ without the additional Adam optimization.
We used our studies on single-spot focusing and determined an optimal value for the learning rate using the bisection method.
The resulting data were fitted with $\mathrm{LR}=aM^b+c$.
For transmission-matrix optimization, the fitted learning rates were reduced by a constant scaling factor.

\section*{Transmission loss of the target modes}

To quantify the optical loss of the implemented transformation, we consider the $N$ target modes where all optical power is being coupled to at the MMW input. 
As such, our metric accounts for all optical power that
does not remain within the $N$ output modes.
Let $\mathbf{TM}_N$ denote the $N\times N$ submatrix of the full $N^o\times N^o$ transmission operator that may also contain additional ancilla modes, with $N^o\ge N$.
The average fraction of input power retained within the target subspace is defined as
\begin{equation}
    P_\eta
    =
    \frac{1}{N}\left\|\mathbf{TM}_N\right\|_{\mathrm{F}}^2
    =
    \frac{1}{N}\sum_{i=1}^{N}\sigma_i^2,
\end{equation}
where $\|\cdot\|_{\mathrm{F}}$ denotes the Frobenius norm and $\sigma_i$ are the singular values of $\mathbf{TM}_N$.
Equivalently, $P_\eta$ corresponds to the output power
contained in the $N$ target modes, averaged over all target modes.
We define the target-subspace loss as
\begin{equation}
    L_{\eta}
    =
    -10\log_{10}\!\left(P_\eta\right).
\end{equation}
This metric treats all power transmitted outside the $N$-dimensional modal space as loss. It therefore includes both coupling to higher-order guided modes outside the target space and power coupled to radiation modes that is no longer guided by the waveguide.

\section*{Thermal perturbation-induced modal coupling}

\begin{figure}[h]
    \centering
    (a)\includegraphics[height=4.2cm]{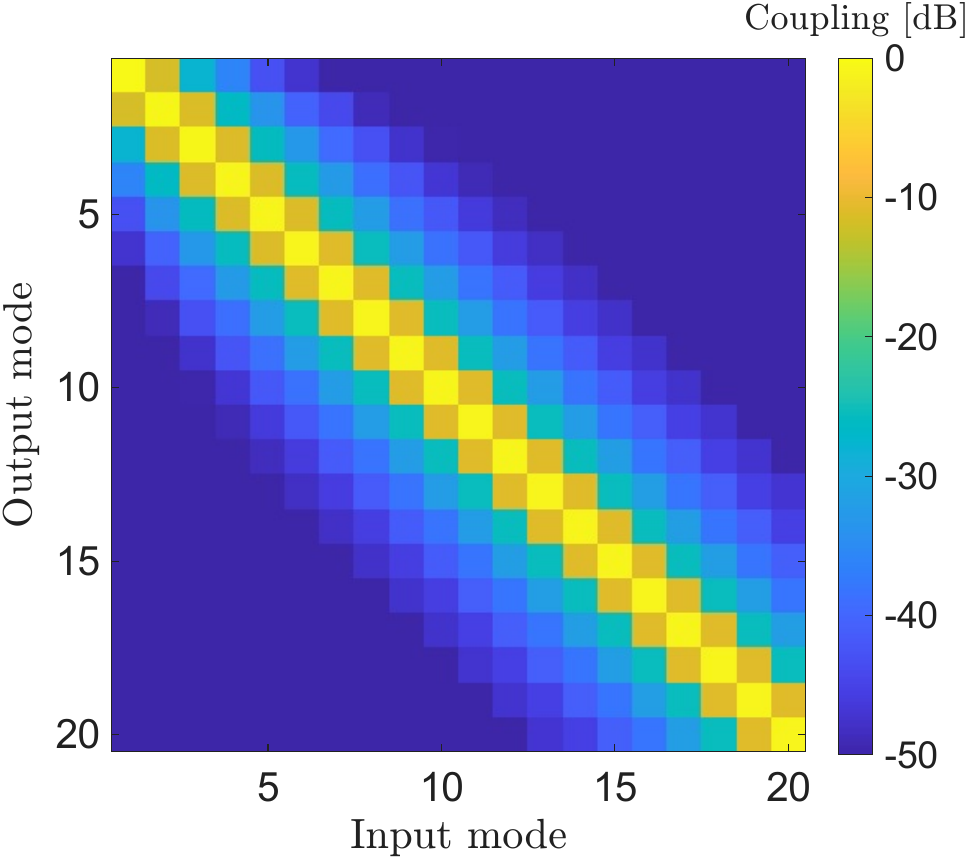}
    (b)\includegraphics[height=4.2cm]{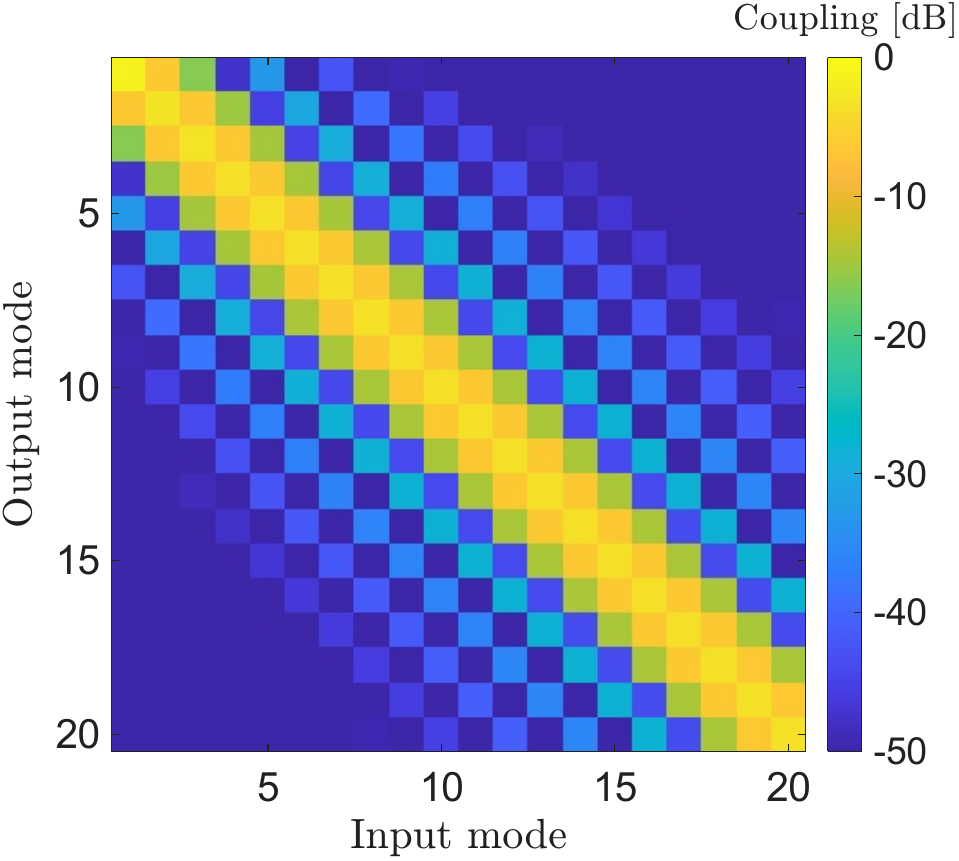}
    (c)\includegraphics[height=4.2cm]{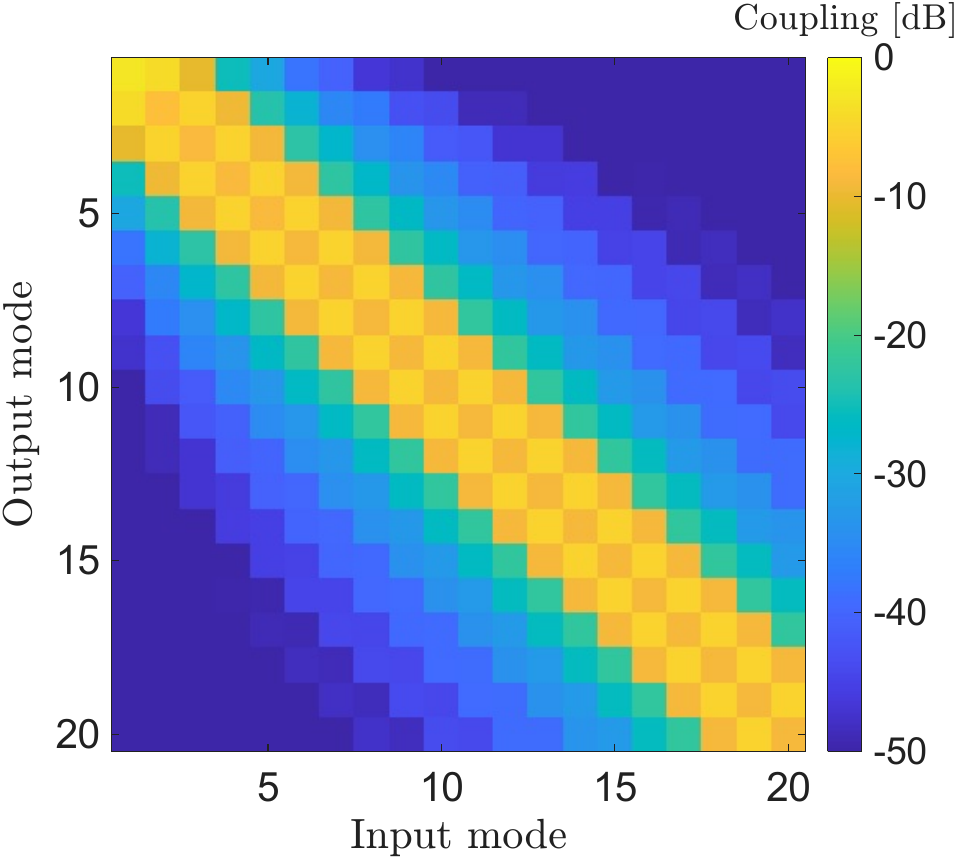}
    \caption{{Modal coupling induced by a spatial phase gradient.}
    Coupling matrices $C_{ij}$ among the first 20 guided modes after application of a spatial phase perturbation with gradient slopes
    (a) $\xi=0.25$, (b) $\xi=0.50$, and (c) $\xi=0.75$.
    Increasing the phase-gradient slope enhances the coupling between the guided modes, with the strongest coupling occurring predominantly between modes with neighboring mode indices.}
    \label{fig:thermalCoupling}
\end{figure}

Spatial phase perturbation modifies the transverse field distribution and thereby couples the otherwise orthogonal guided modes.
To study the effect of this coupling, we perturb the spatial phase of each mode $E_i(x,y)$ individually.
The effect of the thermo-optic perturbation was modeled in MATLAB as a linear phase gradient across the transverse $x$-direction of the MMW. 
For each spatial sampling point $x$, the applied phase was calculated as

\begin{equation}
    \Delta\phi(x;\xi) 
    = 
    \xi\,\left(\frac{2\pi}{W}x - 2\pi\Theta(-\xi) \right)+\phi_0, \;\;\;
    \Theta(\xi) = 
    \begin{cases}
    1 & \text{if } \xi > 0 \\
    0 & \text{if } \xi < 0
\end{cases}
\end{equation}

where $W$ denotes the width of the MMW and $\xi$ is a dimensionless parameter that controls the slope of the phase gradient. A conditional offset $\phi_0$ is included through the Heaviside function in order to ensure positive phase perturbations only.
The corresponding perturbed field was obtained by point-wise multiplication of the complex field with the phase factor
\begin{equation}
    E_{i,\mathrm{pert}}(x,y)
    =
    E_i(x,y) e^{i\Delta\phi(x,y;\xi)},
\end{equation}
and its overlap with each unperturbed mode $E_j(x,y)$ is evaluated using
\begin{equation}
C_{ij}(\xi)
=
\frac{
\left|
\iint E_{i,\mathrm{pert}}(x,y)
E_j^*(x,y)\,\mathrm{d}x\,\mathrm{d}y
\right|^2
}{
\left(\iint |E_i(x,y)|^2\,\mathrm{d}x\,\mathrm{d}y\right)
\left(\iint |E_j(x,y)|^2\,\mathrm{d}x\,\mathrm{d}y\right)
}.
\end{equation}
Figure~\ref{fig:thermalCoupling} shows the resulting coupling matrices among the first 20 guided modes for phase-gradient slopes of $\xi=0.25$, $0.50$, and $0.75$, respectively.
Increasing the gradient slope results in progressively stronger modal coupling.
The coupling occurs predominantly between modes with neighboring mode indices, while coupling between more widely separated modes remains negligible.

\section*{Optimization of randomly rotated transformations}

\begin{figure}[h]
    \raggedright
    (a)\includegraphics[height=3cm]{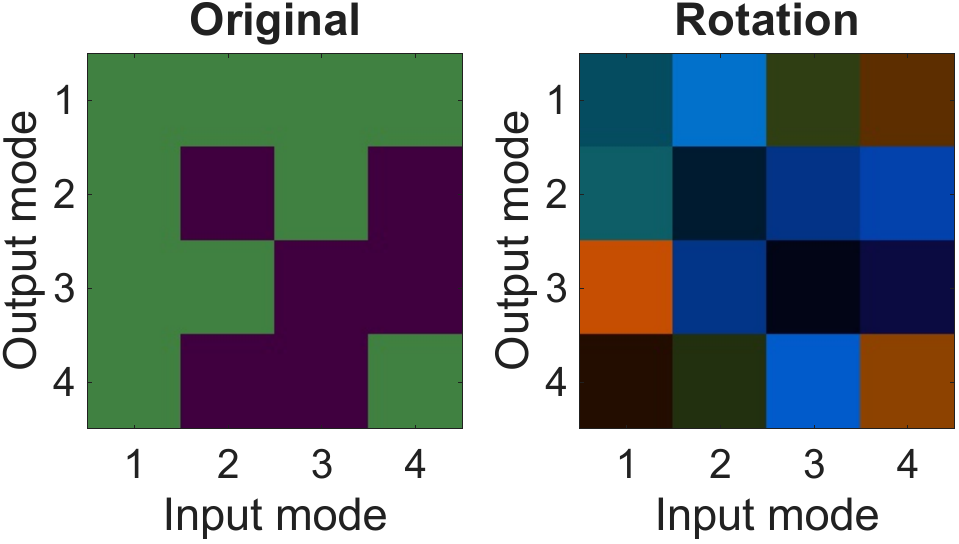}
    (b)\includegraphics[height=4.5cm]{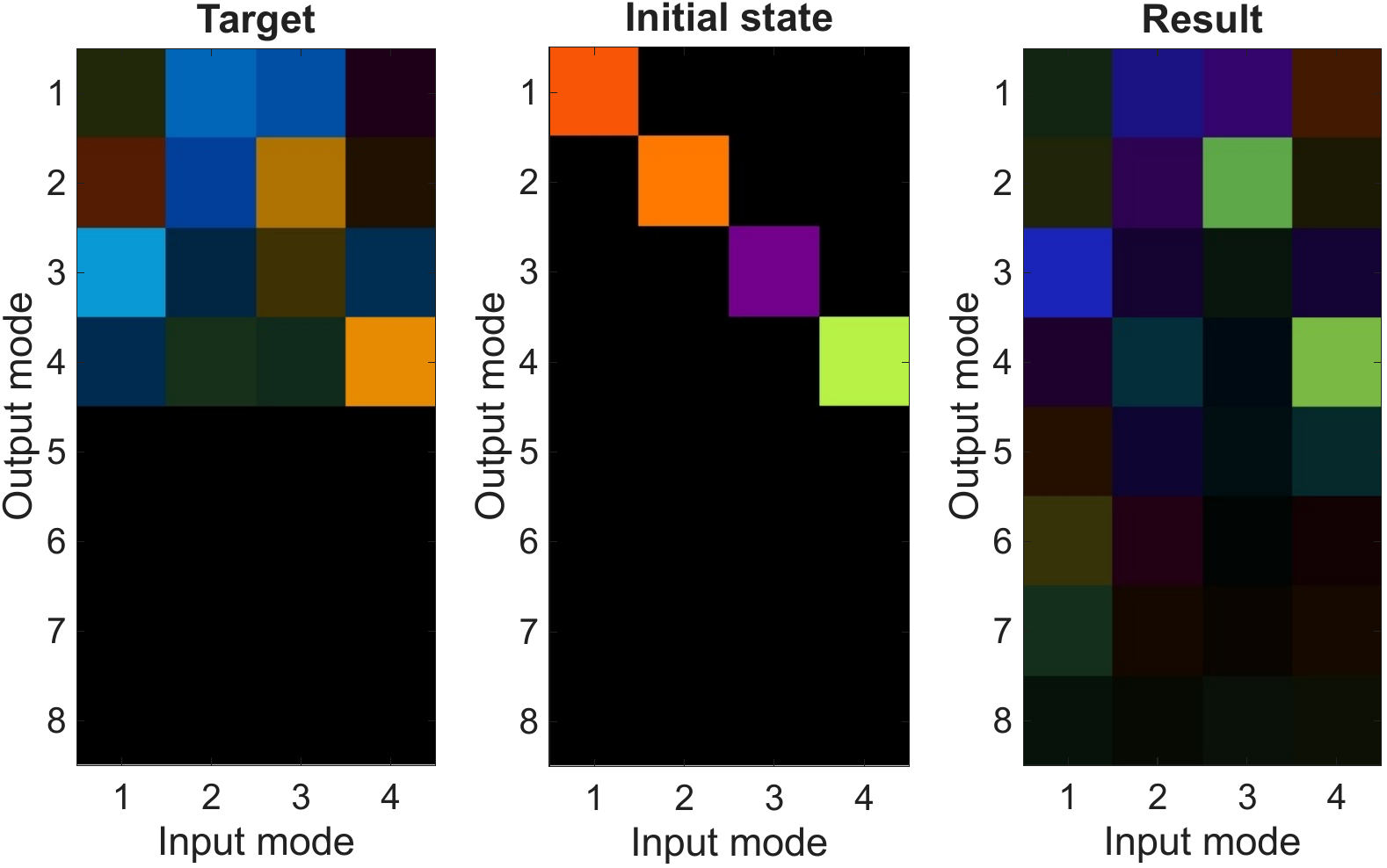}
    \includegraphics[height=4.5cm]{img/complex_colorbar.eps}\\
    \caption{Optimization of a randomly rotated unitary transformation. (a) Original $4\times 4$ Sylvester--Hadamard matrix (left) and randomly generated $4\times 4$ unitary rotation matrix (right). Color encodes the optical phase and brightness the field amplitude. (b) Rotated target transformation, comprising four target and four ancillary modes (left), initial transformation of the unperturbed MMW (center), and optimized transmission matrix after $1.5\times10^5$ iterations (right). The optimization uses $M=5N^2=80$ thermo-optic control elements and reaches a correlation of $98.3\%$ with the target transformation at an optical loss of $L_\eta=1.0$~dB.}    
\label{fig:rotatedTransformation}
\end{figure}

We also investigate the realization of transformations other than the Sylvester--Hadamard matrices considered in the main text.
We consider a randomly rotated unitary transformation. 
We generate a random $4\times4$ unitary matrix and apply it to the $4\times4$ Sylvester--Hadamard matrix, yielding the target transformation.
Since both matrices are unitary, the resulting target transformation remains unitary.
The original Sylvester--Hadamard matrix and the random unitary rotation are shown in the left and right panels of Fig.~\ref{fig:rotatedTransformation}a, respectively.

The resulting rotated transformation, including four ancilla modes, is shown in the left panel of Fig.~\ref{fig:rotatedTransformation}. The optimization is performed using $M=5N^2=80$ thermo-optic control elements and four ancillary modes. The minimum output transmission is set to $\eta_{\mathrm{out,min}}=-1$~dB.
All thermo-optic perturbations are initialized to zero, resulting in the initial transmission matrix shown in the center panel of Fig.~\ref{fig:rotatedTransformation}b. 
After $1.5\times10^5$ optimization iterations, the realized transformation reaches a correlation of $98.3\%$ with the target and with an optical loss of $L_\eta=1.02$~dB. 
The optimization result is shown in the right panel of Fig.~\ref{fig:rotatedTransformation}c.

\section*{Participation ratio of the thermo-optic control elements}

To quantify how the optimized thermo-optic modulation is distributed among the individual control elements, we calculate the participation ratio of the optimized phase-gradient slopes $\boldsymbol{\xi}={\xi_1,\xi_2,\ldots,\xi_M}$. 
We define the participation ratio as

\begin{equation}
\mathrm{\epsilon}
= 
\frac{
\left(\sum_{k=1}^{M} \lvert \xi_k \rvert \right)^2
}{M\cdot
\sum_{k=1}^{M} \lvert \xi_k \rvert ^2
}.
\label{eq:participation_ratio}
\end{equation}

The participation ratio provides a measure of the effective number of thermo-optic control elements contributing to the optimized transformation.
For $\epsilon =\frac{1}{M}$, the modulation is concentrated in a single heater, whereas $\epsilon =1$ corresponds to equal modulation amplitudes across all $M$ heaters. Intermediate values indicate that the required modulation is distributed over a subset of the available control elements.
The participation ratio therefore allows us to assess whether a given optical transformation relies on distributed control throughout the MMW or is dominated by only a few thermo-optic perturbations.

\section*{Influence of the loss term on single-spot focusing}
\begin{figure}[h]
\raggedright
(a)\includegraphics[width=0.46\textwidth]{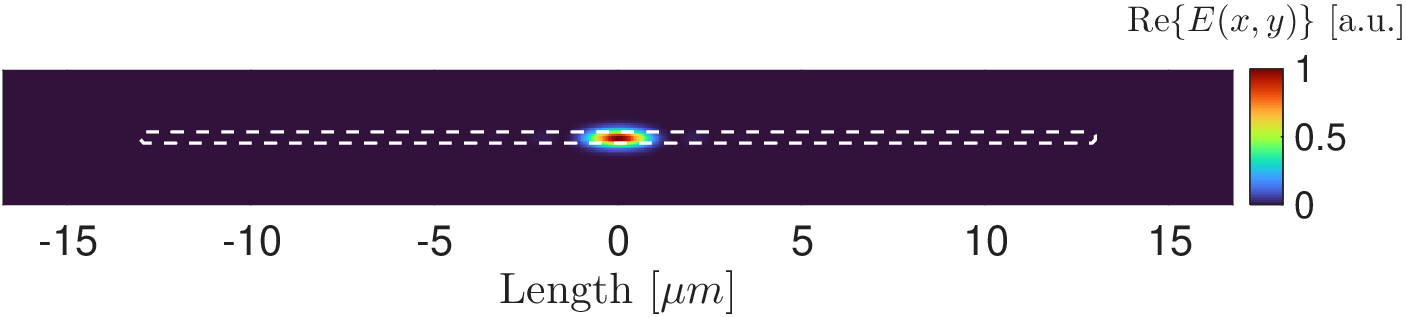}
(b)\includegraphics[width=0.46\textwidth]{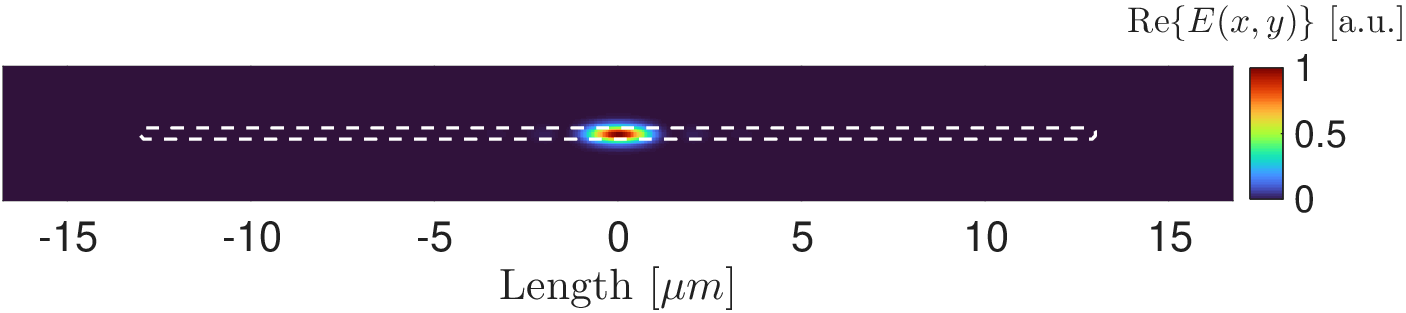}\\
(c)\includegraphics[width=0.46\textwidth]{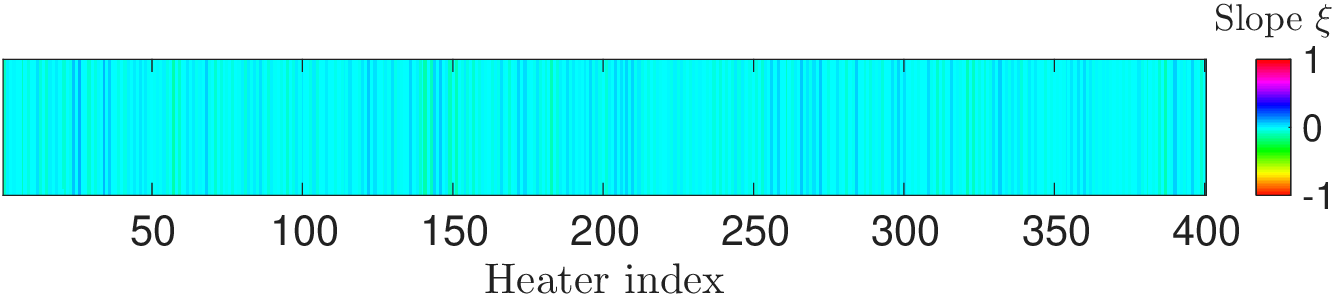}
(d)\includegraphics[width=0.46\textwidth]{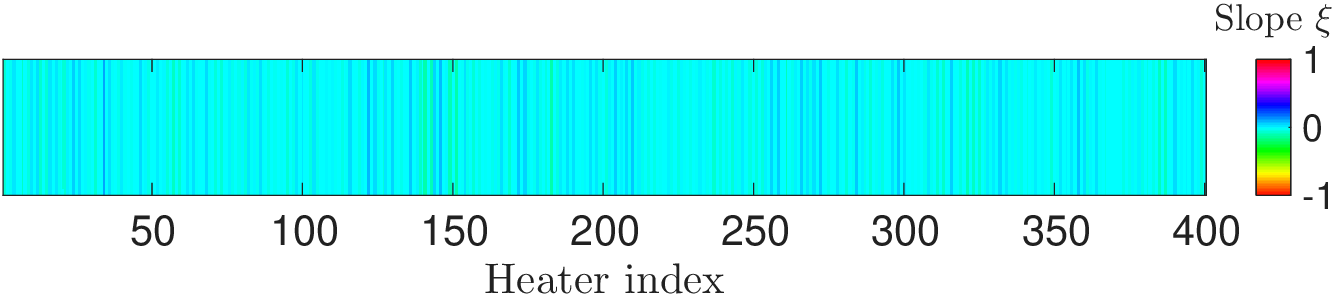}\\
(e)\includegraphics[width=0.4\textwidth]{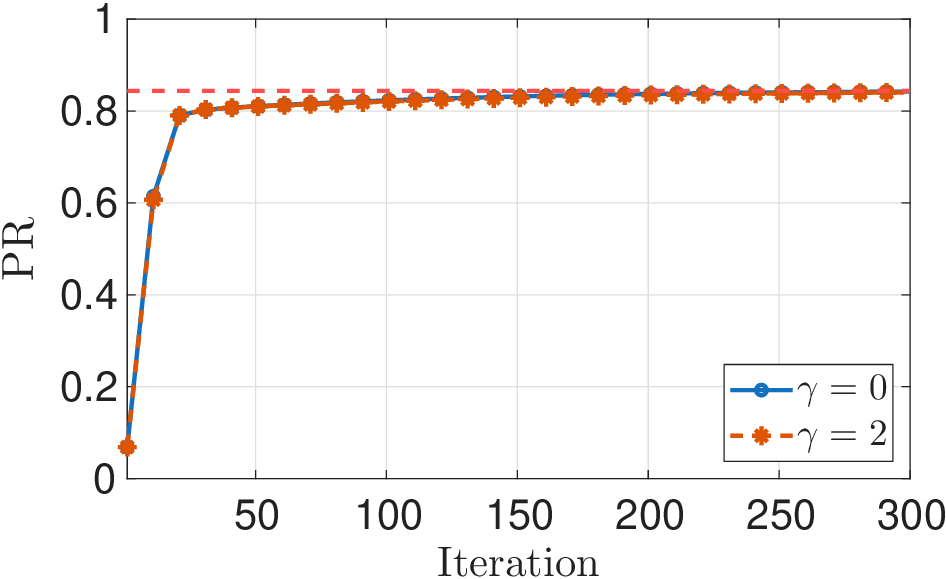}
(f)\includegraphics[width=0.4\textwidth]{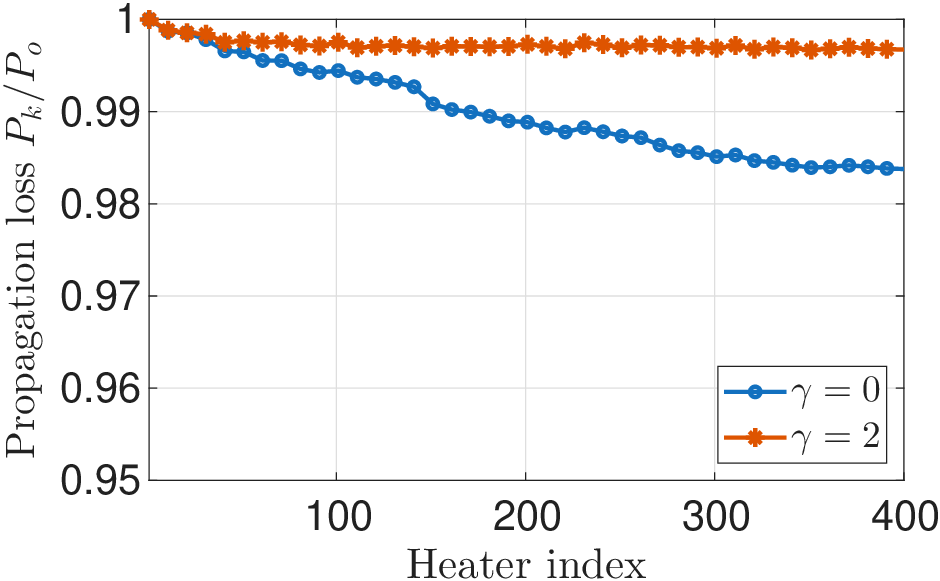}
\caption{
Influence of the transmission weighting $\gamma$ in the cost function Eq.~(\ref{eq:cost_single_spot_focusing}) on single-spot focusing. Both optimizations start from the same initial modal excitation and output field shown in Fig.~\ref{fig:3}a.
(a,b) Optimized output intensity distributions for (a) $\gamma=0$ and (b) $\gamma=2$.
(c,d) Corresponding optimized phase-gradient slopes $\boldsymbol{\xi}$ for (c) $\gamma=0$ and (d) $\gamma=2$. The participation ratios $\eta$ are 161 and 155, respectively.
(e) Evolution of the power ratio during optimization for both values of $\gamma$, showing nearly identical focusing performance.
(f) Normalized forward-propagating optical power along the optimized MMW. Including the transmission term with $\gamma=2$ reduces perturbation-induced optical loss.
}
\label{fig:gamma_comparison}
\end{figure}

The cost function for single-spot focusing introduced in Eq.~(\ref{eq:cost_single_spot_focusing}) contains the normalized output power weighted by the parameter $\gamma$, allowing propagation loss to be explicitly considered during optimization. While the results presented in Fig.~\ref{fig:3} use $\gamma=0$, here we investigate the influence of this term by comparing optimizations with $\gamma=0$ and $\gamma=2$. Both optimizations start from the same modal excitation and heater settings and therefore from the identical initial output intensity distribution shown in Fig.~\ref{fig:3}a.

Figure~\ref{fig:gamma_comparison} compares the resulting optimizations for both values of $\gamma$. In both cases, the optimization produces a sharply confined focal spot within the target region, with nearly identical final power ratios and PR evolution. The corresponding phase-gradient distributions are also very similar, with participation ratios of $\eta=161$ and $155$ for $\gamma=0$ and $\gamma=2$, respectively, out of a total of $M=400$ thermo-optic control elements. Thus, including the transmission term only weakly affects the number of heaters effectively participating in the optimized transformation.

In contrast, the influence of $\gamma$ is clearly visible in the retained optical power. For $\gamma=0$, $98.38\%$ of the input power remains after the final heater, corresponding to a propagation loss of $0.071$~dB. For $\gamma=2$, the retained power increases to $99.67\%$, reducing the loss to only $0.014$~dB. Thus, including the transmission term reduces the perturbation-induced loss by approximately a factor of five while preserving both the achievable focusing performance and the overall participation of the heater array.

\section*{Supplementary algorithm}

\begin{algorithm}[h]
\caption{Gradient-based optimization of the programmable multimode waveguide}
\label{alg:mmw_optimization}
\begin{algorithmic}[1]

\Require
Target transformation $\mathbf{TM}_{\mathrm{target}}\in\mathbb{C}^{N\times N}$;
guided mode profiles $\{\psi_j\}_{j=1}^{N}$;
number of control elements $M$;
control bounds $\mathbf{u}_{\min},\mathbf{u}_{\max}$;
learning rate $\alpha$;
number of iterations $S$;
Modal target space $N$;
transmission threshold $P_{\min}$;
target-power threshold $P_{\mathrm{t,min}}$;
weight $\gamma_{\mathrm{t}}$.

\State Initialize control parameters
$\boldsymbol{\xi}\leftarrow\boldsymbol{\xi}_0$

\State Initialize Adam optimizer

\For{$s=1,\ldots,S$}
\State Generate random-phase input and assign to input modal vector $\mathbf{a}_{\mathrm{in}} \leftarrow [a_1,...,a_N]$
   \[
    a_n =
    \begin{cases}
    \exp(i\varphi_n), & n\leq N,\\
    0, & n>N,
    \end{cases}
    \qquad
    \varphi_n\sim\mathcal{U}(0,2\pi)
    \]

    \State Compute target modal vector
    $\mathbf{\hat{y}}
    \leftarrow \mathbf{TM}_{\mathrm{target}}\mathbf{a}_{\mathrm{in}}$

    \State Propagate $\mathbf{b}
    \leftarrow
    \mathrm{Forward}    (\mathbf{a}_{\mathrm{in}},\boldsymbol{\xi},\{\psi_n\})$
    
    \Comment{Following Eq.~(\ref{eq:perturbation}):
    $E_{\mathrm{pert}}(z)=E(z)e^{i\Delta \phi(\xi)}$,
    $E(z+1)=E_{\mathrm{pert}}(z)e^{i\beta \Delta z}$}

    \State Compute total transmitted power  $P_{\mathrm{out}}
    \leftarrow
    \frac{\|\mathbf{b}\|_2^2}{\|\mathbf{a}_{\mathrm{in}}\|_2^2}$

    \State Restrict output to the target-modes
    $\mathbf{{b}_T}
    \leftarrow
    [b_1,\ldots,b_{N}]^{\mathrm{T}}$
    
    \State Compute fraction of power in the target modes
    $P_{\mathrm{target}}
    \leftarrow
    \frac{\|\mathbf{b_T}\|_2^2}
         {\|\mathbf{b}\|_2^2}$
    
    \State Compute transformation fidelity
    $
    F\leftarrow
    \frac{
    \left|
    \mathbf{\hat{y}}^{\dagger}
    \mathbf{b}
    \right|^2}
    {
    \|\mathbf{\hat{y}}\|_2^2
    \|\mathbf{b}\|_2^2
    }$

    \State Compute loss terms
    $    \mathcal{L}_1\leftarrow1-F,
    \mathcal{L}_2\leftarrow
    \max(0,P_{\min}-P_{\mathrm{out}}),
    \mathcal{L}_3\leftarrow
    \gamma_{\mathrm{t}}
    \max(0,P_{\mathrm{t,min}}-P_{\mathrm{t}})$

    \State Compute total loss
    $\mathcal{L}\leftarrow \mathcal{L}_1+\mathcal{L}_2+\mathcal{L}_3$

    \State Compute gradient by automatic differentiation
    $\mathbf{g}_s
    \leftarrow \nabla_{\boldsymbol{\xi}}\mathcal{L}$

    \State Update phase-gradient slopes using Adam~\cite{kingma2014adam}
    $\boldsymbol{\xi}^+
    \leftarrow
    \mathrm{Adam}(\boldsymbol{\xi},\mathbf{g},\alpha)
    $

    \State     {Constrain phase-gradient slopes $\boldsymbol{\xi}^+
    \leftarrow \mathrm{Clip}(\boldsymbol{\xi}^+, \mathbf{u}_{\min},\mathbf{u}_{\max})$
    }

   \If{$s \bmod 1000 = 0$}
        \State Instantaneous transformation
        $
        \mathbf{TM}_{\mathrm{MMW}}
        \leftarrow
        \mathrm{TransmissionMatrix}(\boldsymbol{\xi})
        $
        \State Evaluate matrix correlation with the target transformation
        $c
        \leftarrow
        \frac{
        \left|
        \mathrm{Tr}
        \left(
        \mathbf{TM}_{\mathrm{target}}^{\dagger}
        \mathbf{TM}_{\mathrm{MMW}}
        \right)
        \right|
        }
        {
        \|\mathbf{TM}_{\mathrm{target}}\|_{\mathrm{F}}
        \|\mathbf{TM}_{\mathrm{MMW}}\|_{\mathrm{F}}
        }$

        \If{$c > c_{\mathrm{previous}}$}
            \State 
            {
            Keep current best phase-gradient slopes $\boldsymbol{\xi_{\mathrm{best}}} \leftarrow \boldsymbol{\xi}$
            }
            
            \State 
            {
            Keep current best instantaneous transformation $\mathbf{TM}_{\mathrm{MMW, best}} \leftarrow \mathbf{TM}_{\mathrm{MMW}}$
            }

            \State 
            {
            Update previous best correlation value  $c_{\mathrm{previous}} \leftarrow c$
            }
        \EndIf 
        
    \EndIf
    
\EndFor


\State \Return {
optimized phase-gradient slopes
$\boldsymbol{\xi_{\mathrm{best}}}$ and resulting transformation
$\mathbf{TM}_{\mathrm{MMW, best}}$
}

\end{algorithmic}
\end{algorithm}

\end{appendices}

\bibliographystyle{IEEEtran}
\bibliography{modernize-literature}

\end{document}